%% file: paper.tex
\documentclass{article}
\usepackage{spconf,amsmath,graphicx}
\usepackage{bm}
\usepackage{pgfplots}
\usepackage{flushend}
\usepackage[ruled,vlined]{algorithm2e}
\usepackage{tikz}
\usepackage{multirow}
\usepackage{hyperref}
\DeclareUnicodeCharacter{2212}{−}
\usepgfplotslibrary{groupplots,dateplot}
\usetikzlibrary{patterns,shapes,arrows,arrows.meta}
\pgfplotsset{compat=newest}
\usetikzlibrary{positioning}
\graphicspath{{figures/}}

\let\OLDthebibliography\thebibliography
\renewcommand\thebibliography[1]{
	\OLDthebibliography{#1}
	\setlength{\parskip}{0pt}
	\setlength{\itemsep}{5pt plus 0.5ex}
}

\usepackage[labelsep=period]{caption}
\usepackage{textcomp}
\renewcommand{\thetable}{\Roman{table}}

\newcommand{\fig}{Fig.\ }
\newcommand{\tab}{Table }

\newcommand{\channel}{c}
\newcommand{\x}{x}
\newcommand{\y}{y}
\newcommand{\z}{z}

\newcommand{\scalex}{s_x}
\newcommand{\scaley}{s_y}
\newcommand{\scalez}{s_z}

\newcommand{\referenceimage}{R}
\newcommand{\networkoutput}{O}
\newcommand{\msmask}{M}
\newcommand{\msmasked}{D}
\newcommand{\msimage}{I}

\newcommand{\gridwidth}{G_w}
\newcommand{\gridheight}{G_h}
\newcommand{\gridbins}{G_l}
\newcommand{\lowLRA}{A^l}
\newcommand{\lowLRB}{B^l}
\newcommand{\highLRA}{A^h}
\newcommand{\highLRB}{B^h}
\newcommand{\highwidth}{W}
\newcommand{\highheight}{H}

\AtBeginShipout{\AtBeginShipoutUpperLeft{%
		\put(\dimexpr\paperwidth-20cm\relax,-0.5cm){\parbox[t]{19cm}{\copyright 2023 IEEE. Published in 2023 IEEE International Conference on Image Processing (ICIP), scheduled for 8-11 October 2023 in Kuala Lumpur, Malaysia. Personal use of this material is permitted. However, permission to reprint/republish this material for advertising or promotional purposes or for creating new collective works for resale or redistribution to servers or lists, or to reuse any copyrighted component of this work in other works, must be obtained from the IEEE. DOI: \url{https://doi.org/10.1109/ICIP49359.2023.10222159}}}%
}}

\title{Cross Spectral Image Reconstruction Using A Deep Guided Neural Network}
\name{Frank Sippel, Jürgen Seiler, and André Kaup}
\address{Friedrich-Alexander-Universität Erlangen-Nürnberg\\
	Multimedia Communications and Signal Processing\\
	Cauerstraße 7, 91058 Erlangen, Germany\\
\thanks{The authors gratefully acknowledge that this work has been supported by
the Deutsche Forschungsgemeinschaft (DFG, German Research Foundation) under project number 491814627. \\
\noindent\hspace*{5mm}%
Source code: \textit{\url{https://github.com/FAU-LMS/dgnet}}.
}}

\begin{document}
%
\maketitle

\begin{abstract}
Cross spectral camera arrays, where each camera records different spectral content, are becoming increasingly popular for RGB, multispectral and hyperspectral imaging, since they are capable of a high resolution in every dimension using off-the-shelf hardware.
For these, it is necessary to build an image processing pipeline to calculate a consistent image data cube, i.e., it should look like as if every camera records the scene from the center camera.
Since the cameras record the scene from a different angle, this pipeline needs a reconstruction component for pixels that are not visible to peripheral cameras.
For that, a novel deep guided neural network (DGNet) is presented.
Since only little cross spectral data is available for training, this neural network is highly regularized.
Furthermore, a new data augmentation process is introduced to generate the cross spectral content.
On synthetic and real multispectral camera array data, the proposed network outperforms the state of the art by up to 2 dB in terms of PSNR on average.
Besides, DGNet also tops its best competitor in terms of SSIM as well as in runtime by a factor of nearly 12.
Moreover, a qualitative evaluation reveals visually more appealing results for real camera array data.
\end{abstract}
\section{Introduction}
\label{sec:intro}


Multispectral camera arrays\cite{cam_array_wheat_2021, cam_array_notch_2022} and hyperspectral camera arrays\cite{cam_array_art_2022} are capable of recording different spectral areas by employing multiple cameras.
In contrast to other multispectral and hyperspectral imaging devices, this approach has the advantage of yielding a high resolution in spatial and temporal dimension as well as being cost-efficient and flexible regarding the used filters.
The recorded channels do not necessarily have to lie in the visible wavelength area, but can also contain information about areas in the infrared or ultraviolet area of the spectrum.
These areas of the spectrum are particularly interesting for classification problems.
Multispectral and hyperspectral cameras can be used in medicine to classify the degree of burn \cite{moroni_pet_2015}, in recycling to sort materials \cite{moroni_pet_2015}, in forensics by determining the age of blood \cite{edelman_hyperspectral_2013}, or in agriculture by discriminating between parts of fields that need water and fertilizer and those that are healthy \cite{lima_monitoring_2020}.

\begin{figure}[t]
	\begin{center}
		\begin{tikzpicture}[y=-1cm]
			\node[inner sep=0pt] (camsi) at (-0.5,0) {\includegraphics[width=.12\textwidth]{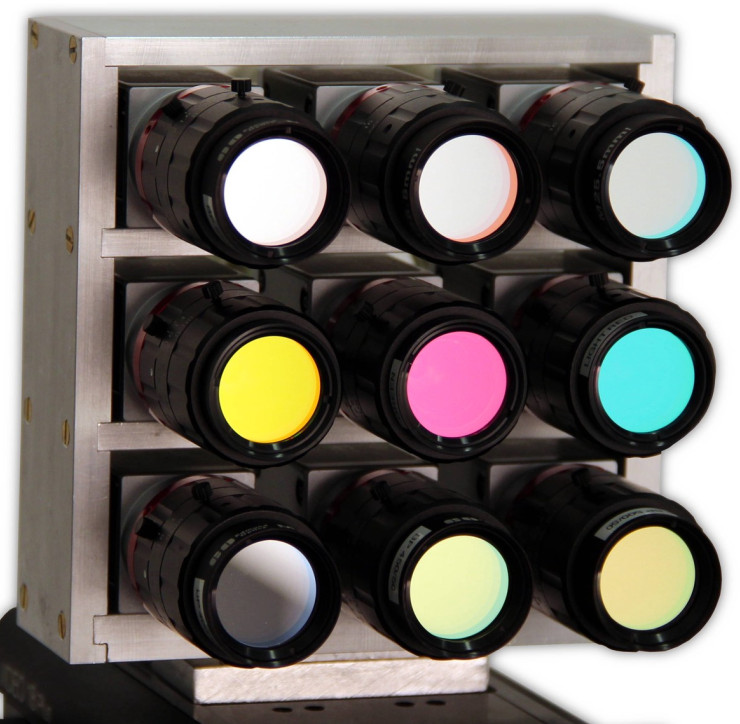}};

			\draw[draw=red, line width=1mm] (-1.2,-0.2) rectangle ++(1.8,0.5);

			\node[align=center] at (2,-0.9) {left\strut};
			\node[align=center] at (3.9,-0.9) {center\strut};
			\node[align=center] at (5.8,-0.9) {right\strut};

			\node[inner sep=0pt] (lef) at (2,0) {\includegraphics[width=.1\textwidth]{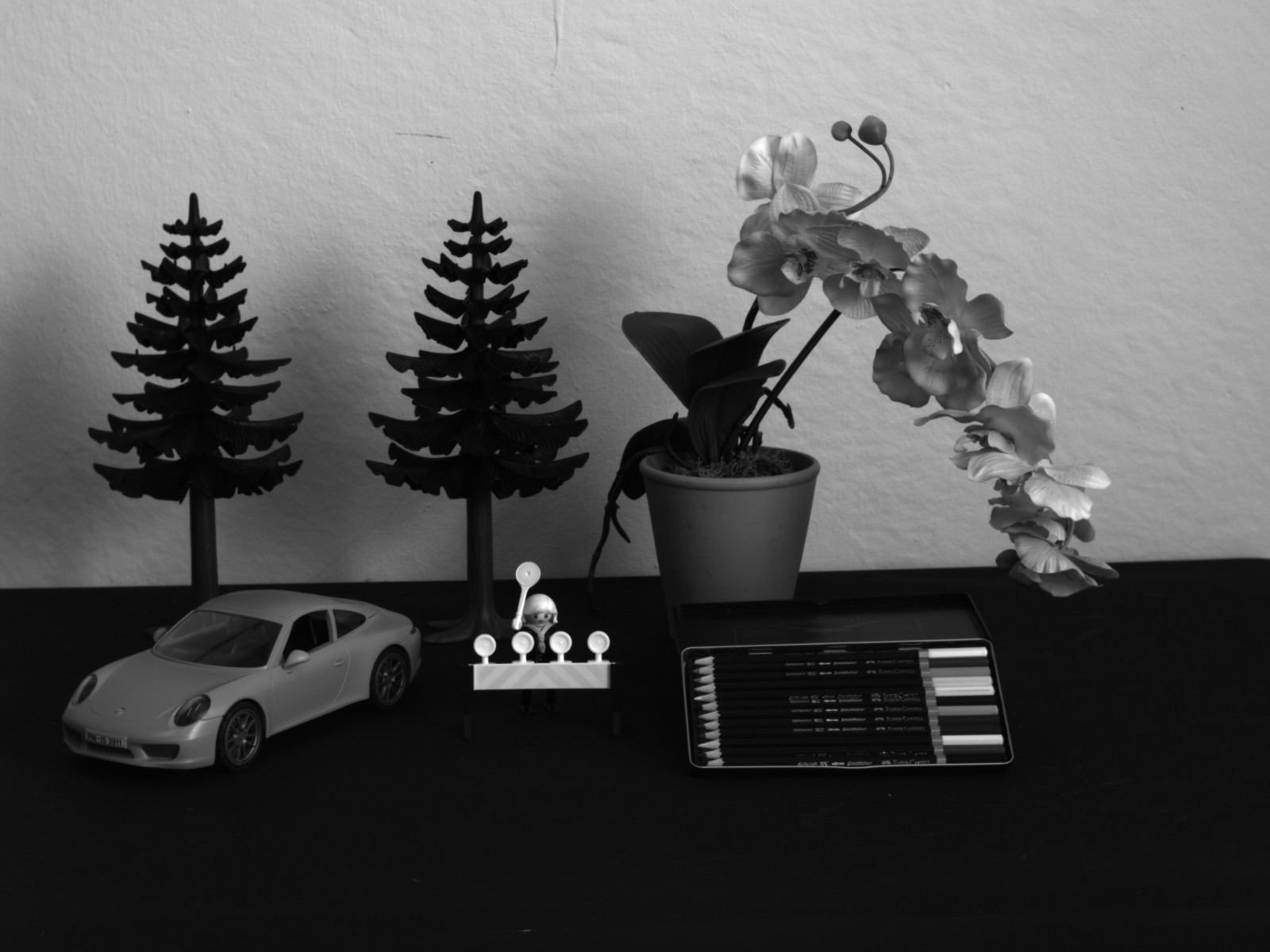}};
			\node[inner sep=0pt] (cen) at (3.9,0) {\includegraphics[width=.1\textwidth]{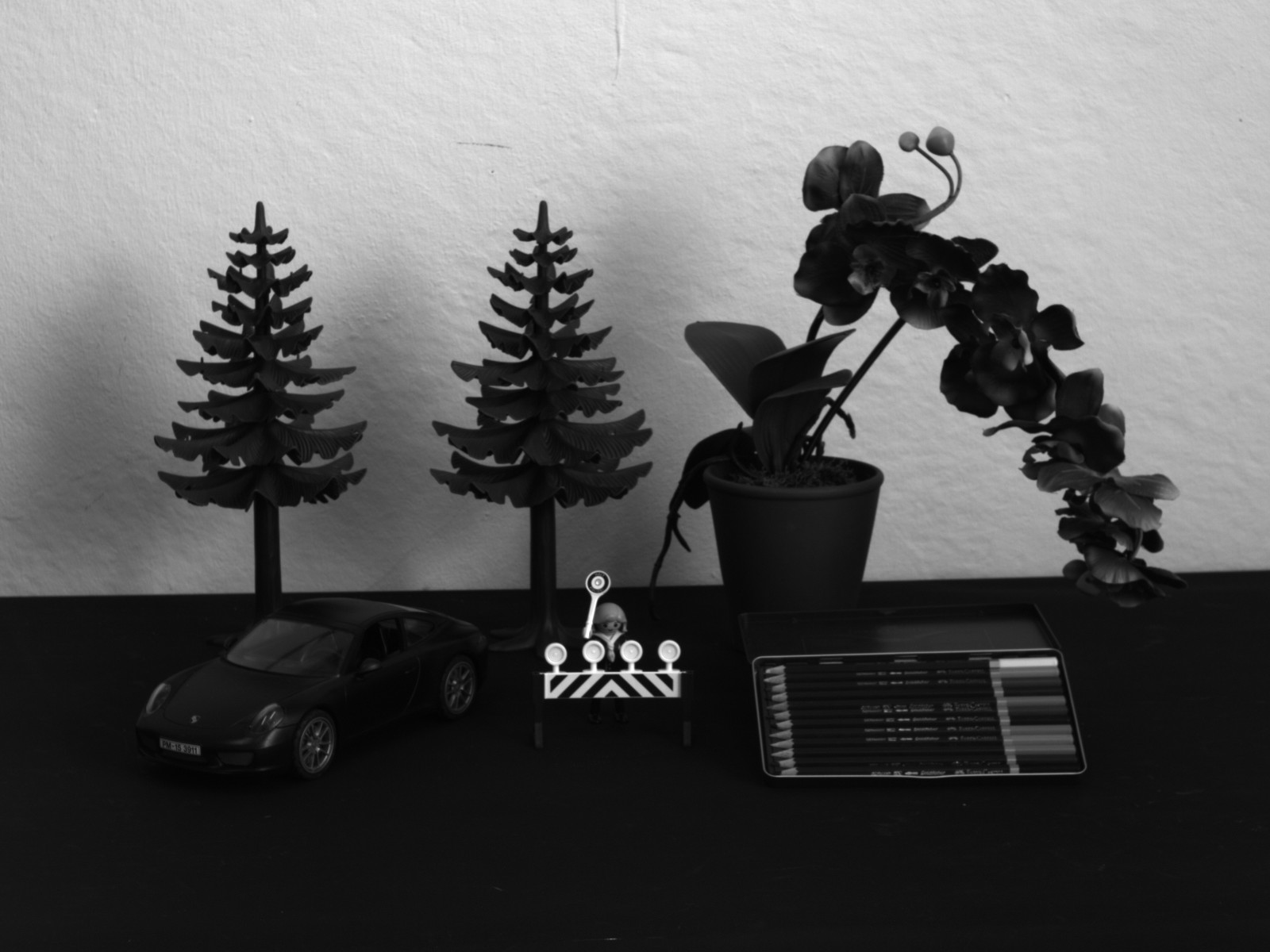}};
			\node[inner sep=0pt] (rig) at (5.8,0) {\includegraphics[width=.1\textwidth]{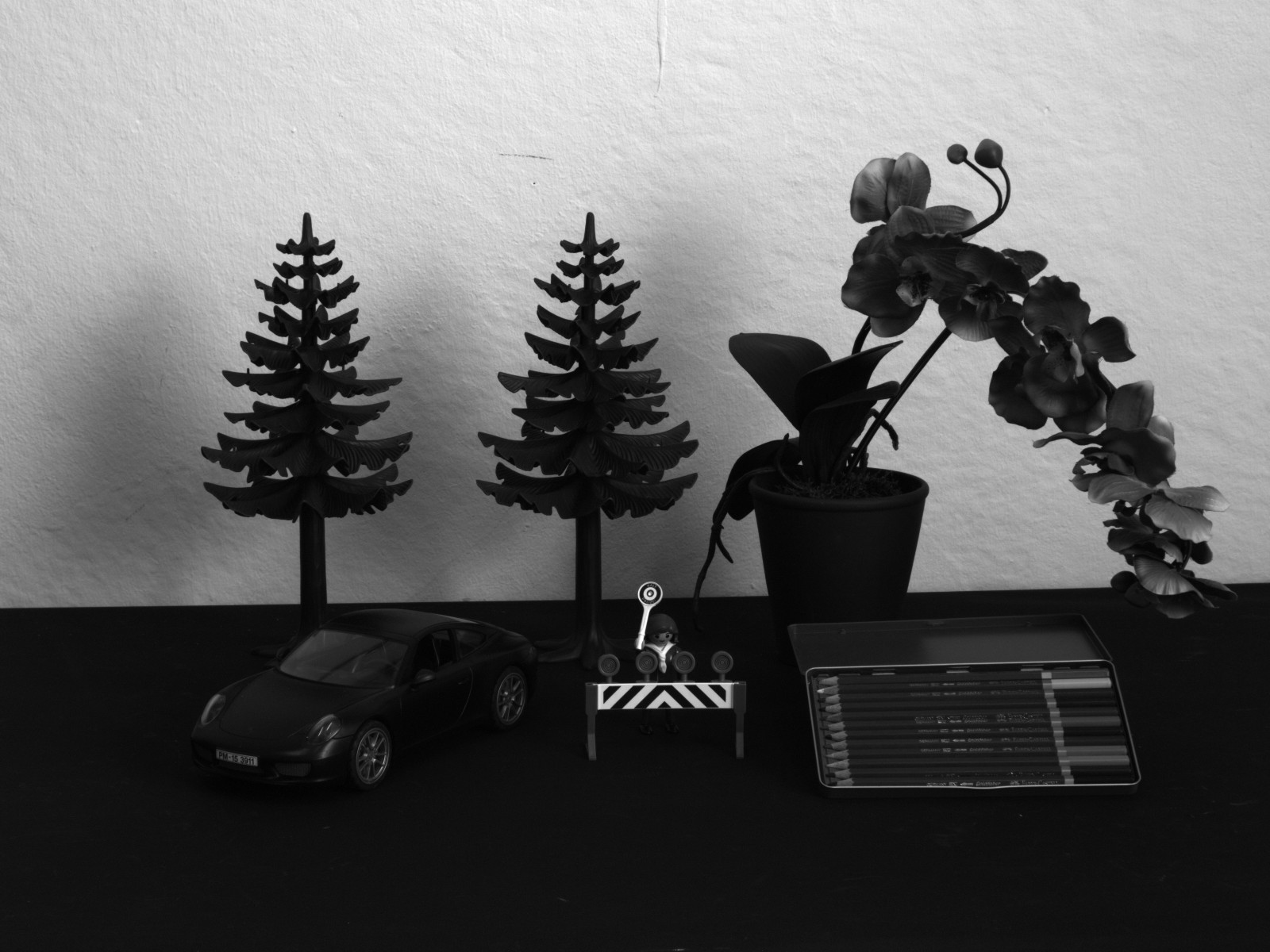}};

			\node[inner sep=0pt] (rgb) at (0,2.7) {\includegraphics[width=.2\textwidth]{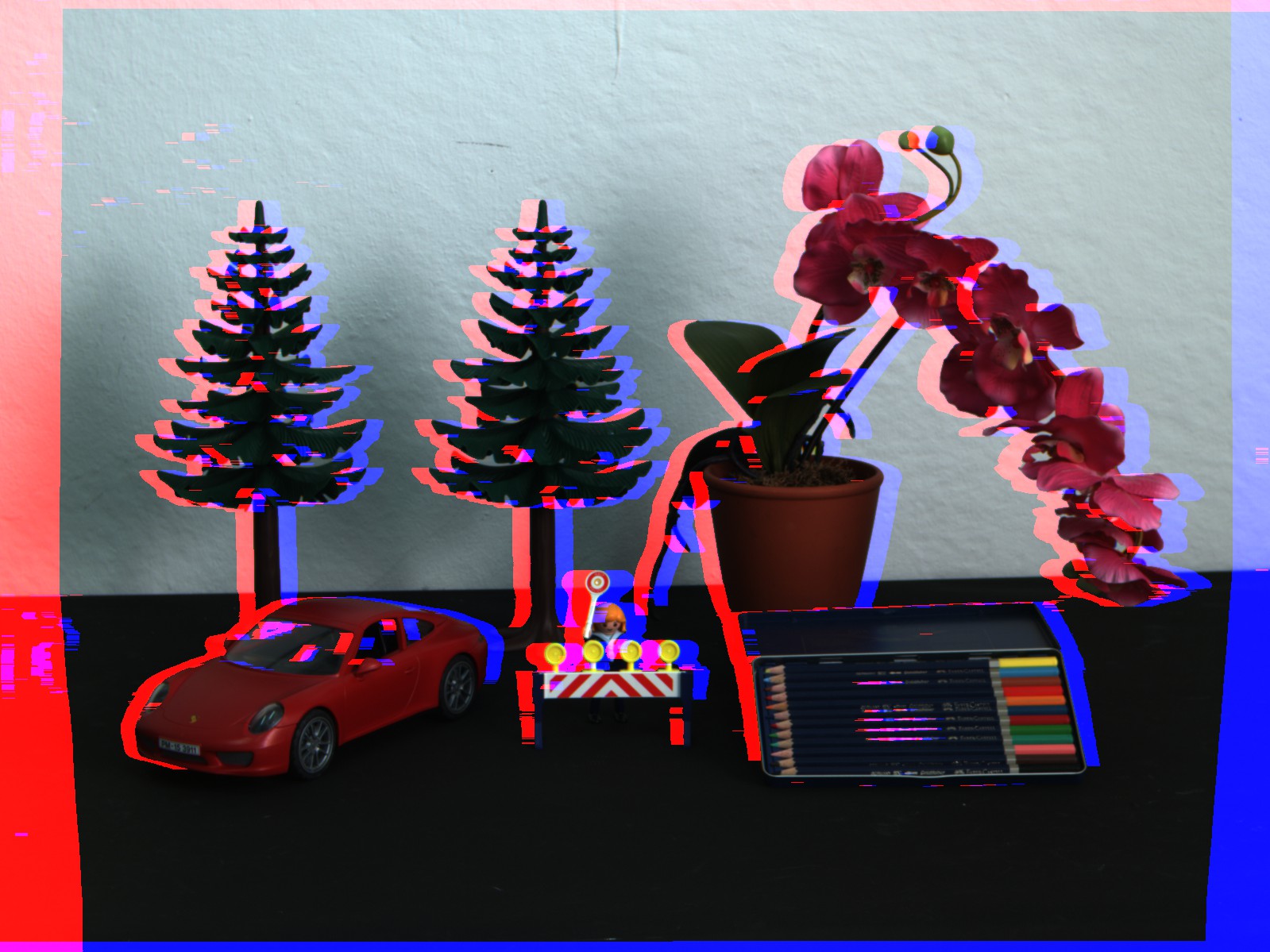}};
			\node[inner sep=0pt] (res) at (4.9,2.7) {\includegraphics[width=.2\textwidth]{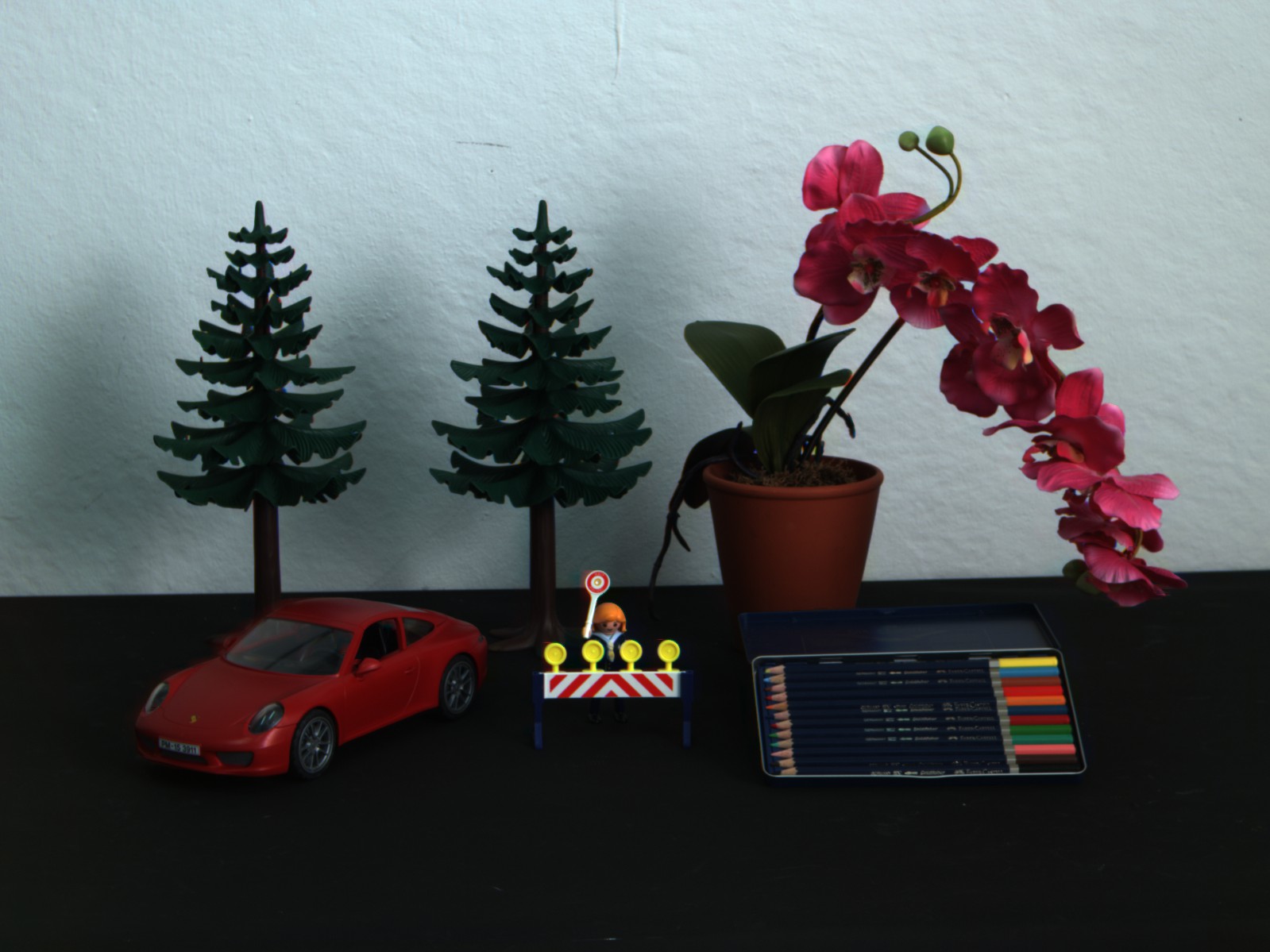}};

			\draw[-{Stealth[scale=2]}] (2, 0.7) -- (0, 1.3);
			\draw[-{Stealth[scale=2]}] (3.9, 0.7) -- (0.9, 1.3);
			\draw[-{Stealth[scale=2]}] (5.8, 0.7) -- (1.8, 1.3);

			\draw[-{Stealth[scale=2]}] (0, 4.2) -- (4.9, 4.2) node[midway, below] (reco) {Reconstruction};
		\end{tikzpicture}
	\end{center}
	\caption{A multispectral camera array (CAMSI) with three corresponding recorded images of the center row (top). Additionally, the registered multispectral image with missing pixels (bottom left) and the reconstructed image (bottom right) are depicted.}
	\label{fig:problem_statement}
\end{figure}
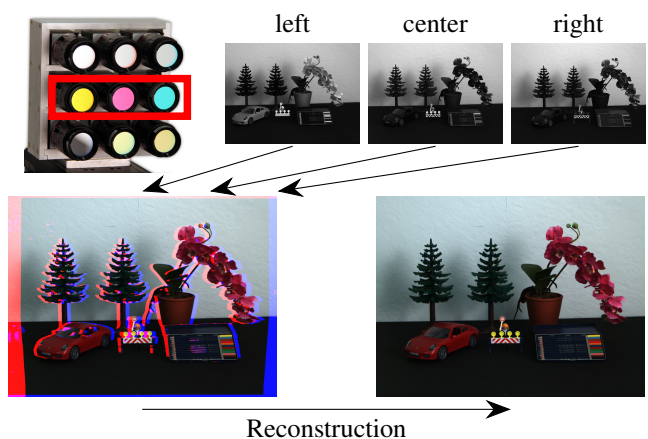

\begin{figure*}[t]
	\begin{center}
		\begin{tikzpicture}[y=-1cm]
			\input{figures/architecture.tex}
		\end{tikzpicture}
	\end{center}
	\caption{The architecture of the proposed network. Blue blocks are convolutional layers with parameters channels/kernel size/stride.}
	\label{fig:network_arch}
\end{figure*}
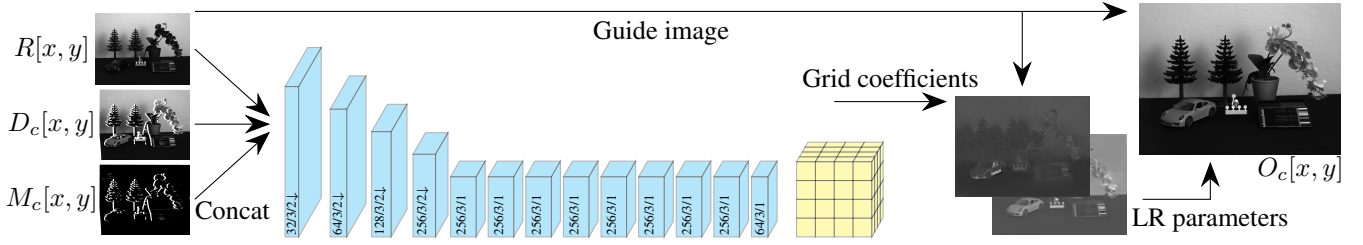

Genser et al. \cite{genser_camera_2020} introduced a multispectral camera array consisting of nine cameras.
However, a reconstruction pipeline to reconstruct a consistent multispectral datacube is necessary.
The goal of this processing pipeline is to reconstruct a consistent multispectral center view by mapping all peripheral cameras to the center perspective, see \fig\ref{fig:problem_statement}.
Then, each pixel of the multispectral datacube shows exactly the same object in different spectral bands.

This reconstruction pipeline consists of multiple separate steps.
First, a calibration procedure is necessary, since the cameras within the camera array as well as the sensors within the camera housing are not perfectly aligned.
For that, a checkerboard pattern is placed in front of the camera, and the images are calibrated to this depth.
Afterwards, since the displacement of the pixels of the peripheral cameras is depth-dependent, a cross spectral depth estimation is performed, e.g., by \cite{genser_deep_2020}.
Subsequently, the pixels of the peripheral cameras can be warped to the center view based on this depth map.
However, due to occlusions the center camera sees pixels that the individual peripheral cameras do not see.
Hence, a cross spectral reconstruction process is necessary that fills in these missing pixels.

Fortunately, this is not a pure inpainting problem as tackled by classical algorithms \cite{getreuer_total_2012, genser_spectral_2018} or neural networks \cite{nazeri_edgeconnect_2019, yu_free_2019, liu_image_2018}, since the complete center view is available as reference image.
However, this reference image lies in a different spectral band, thus only the structure can be used as guiding information for the reconstruction process.
Algorithms from literature that also use this guiding information are \cite{genser_camera_2020} and recently also \cite{sippel_spatio_2021}.
These algorithms are all based on classical image processing techniques like guided filtering and non-local filtering.
However, these methods are not able to establish global relationships between guide pixels and valid pixels in the distorted view.
Furthermore, these methods typically fail in reconstructing high frequency areas of the image, since they are relying on enough similar reference pixel being locally available.
Moreover, until now there are no guided reconstruction methods based on neural networks.
A big challenge of training a neural network for this task is that there is too little multispectral and hyperspectral data available to train such a network properly.

\section{Proposed Network}
\label{sec:network}

The basic idea is to estimate a low dimensional cube of linear regression coefficients through a convolutional neural network.
Afterwards, the cube is sliced into linear regression parameters for each pixel using a guide image.
Finally, these linear regression coefficients are used on the guide image to obtain the final result.

\subsection{Architecture}

In \fig\ref{fig:network_arch}, the network architecture of the proposed Deep Guided Neural Network (DGNet) is depicted.
The input images with pixel coordinates $(\x, \y)$ are the reference image $\referenceimage[\x, \y]$, i.e., the center view, the distorted image $\msmasked_{\channel}[\x, \y]$, i.e., channel $\channel$ of a warped peripheral view, and the mask $\msmask_{\channel}[\x, \y]$, which indicates pixels that need to be reconstructed.
The missing pixels in $\msmasked_{\channel}[\x, \y]$ are set to white as well.
These images are concatenated and put into downscaling convolutional layers with stride 2.
The number of of downscaling layers depends on how many pixels should be within one spatial bin.
For example, for a bin size of $16 \times 16$, 4 downscaling layers are needed.
For every downscaling layer, the number of channels is doubled.
Afterwards, nine convolutional layers without any scaling are applied, which yield cubes of size $\gridwidth \times \gridheight \times (8 \cdot \gridbins)$, where $\gridbins$ is the desired amount of luma bins in the resulting linear regression cube \cite{bilateral_chen_2016, gharbi_deep_2017}.
This also increases the perceptual field, which is necessary for large areas of missing pixels.
Then, a final convolutional layer is applied, which transforms the previous number of channels into the proper number of channels for the linear regression bilateral grid with dimensions $\gridwidth \times \gridheight \times (2 \cdot \gridbins)$.
For this layer, only weights are applied and no activation is used.
Therefore, this layer can be interpreted as a linear combination of features for determining the linear regression coefficients of the grid.
Finally, this tensor is split into two to yield one low-resolution cube for the linear parameter $\lowLRA_{\channel}[\x, \y, \z]$ of size $\gridwidth \times \gridheight \times \gridbins$ and one low-resolution cube for the linear regression bias $\lowLRB_{\channel}[\x, \y, \z]$ of the same size.

The next step is to use the reference image $\referenceimage[\x, \y]$ of size $\highwidth \times \highheight$ to slice the coefficients to two high resolution images of linear regression coefficients $\highLRA_{\channel}$ and $\highLRB_{\channel}$.
For that, trilinear interpolation is used
\begin{equation}
	\highLRA_{\channel}[\x, \y]\hspace{-0.1cm}=\hspace{-0.2cm}\sum_{i, j, k}\hspace{-0.1cm}\theta( \scalex \x - i) \theta( \scaley \y - j) \theta( \scalez \referenceimage[\x, \y] - k) \cdot \lowLRA_{\channel}[i, j, k],
\end{equation}
where $\theta\left(\cdot\right) = \max\left( 1 - | \cdot |, 0 \right)$ is the linear interpolation kernel, $\scalex$ and $\scaley$ are the ratios of the spatial dimensions of the grid with respect to the high-resolution image, i.e., $\scalex = \gridwidth/\highwidth$.
Similarly, $\scalez$ is the ratio of luma bins to the maximum possible intensity value of the grayscale reference image.
Of course, the same interpolation is also performed to yield the high-resolution bias $\highLRB_{\channel}$.
The output image of the network is calculated by applying these coefficients to the reference image
\begin{equation}
	\networkoutput_{\channel}[\x, \y] = \highLRA_{\channel}[\x, \y] \cdot \referenceimage[\x, \y] + \highLRB_{\channel}[\x, \y].
\end{equation}
The resulting multispectral channel $\msimage_{\channel}[\x, \y]$ is calculated by only using the reconstructed image for masked pixels
\begin{equation}
	\msimage_{\channel}[\x, \y] = (1 - \msmask_{\channel}[\x, \y]) \cdot \msmasked_{\channel}[\x, \y] + \msmask_{\channel}[\x, \y] \cdot \networkoutput_{\channel}[\x, \y].
\end{equation}

\subsection{Data Augmentation}

\begin{figure}[t]
	\begin{center}
		\begin{tikzpicture}[y=-1cm]
			\node[inner sep=0pt] (lef) at (0, 0) {\includegraphics[width=.12\textwidth]{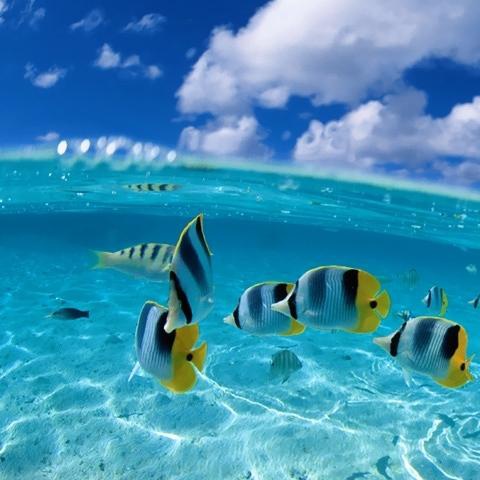}};
			\node[inner sep=0pt] (lef) at (3.8, 0) {\includegraphics[width=.12\textwidth]{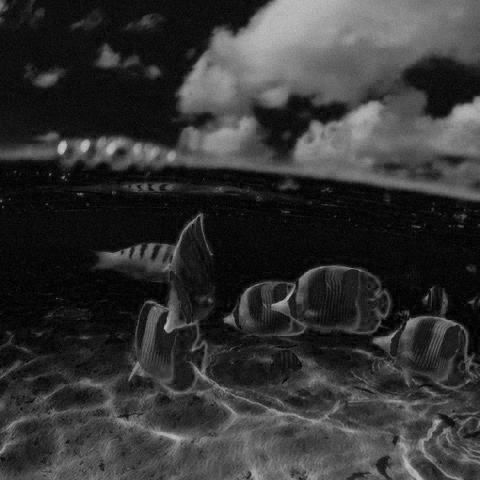}};
			\node[inner sep=0pt] (lef) at (6.2, 0) {\includegraphics[width=.12\textwidth]{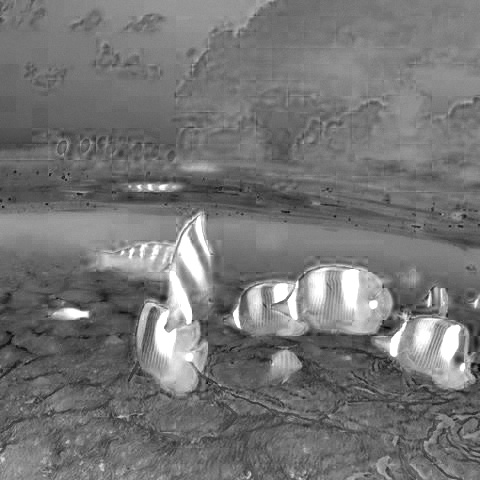}};
			\draw[-{Stealth[scale=2]}] (1.2, 0) -- (2.6, 0);

			\node[inner sep=0pt, draw=black] (lef) at (0.5, 2.5) {\includegraphics[width=.12\textwidth]{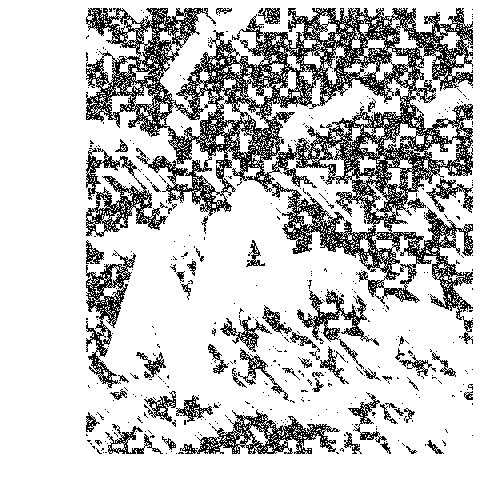}};
			\node[inner sep=0pt, draw=black] (lef) at (3, 2.5) {\includegraphics[width=.12\textwidth]{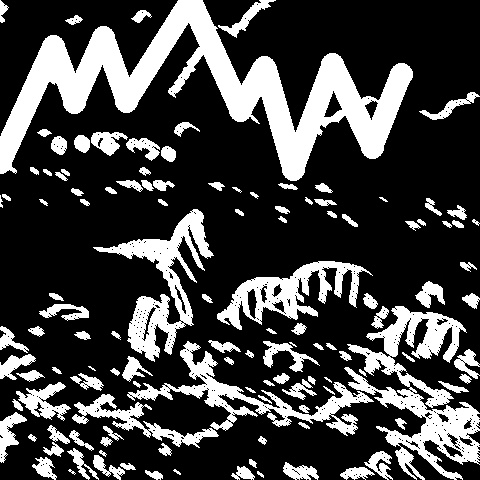}};
			\node[inner sep=0pt, draw=black] (lef) at (5.5, 2.5) {\includegraphics[width=.12\textwidth]{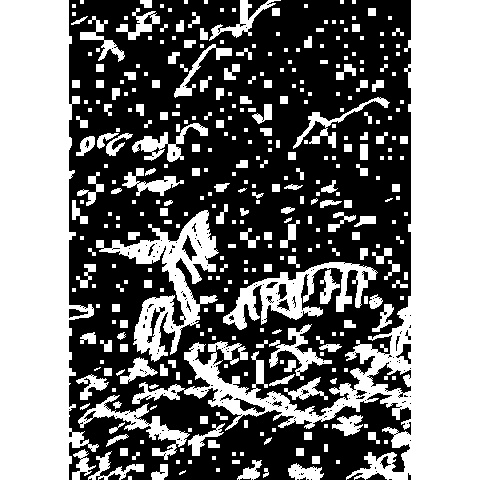}};
		\end{tikzpicture}
	\end{center}
	\caption{In the first row, two generated spectral grayscale images from an RGB image are shown. In the bottom row, three masks are depicted. Note that the edge mask is generated using the RGB image of the first row.}
	\label{fig:data_augmentation}
	\vspace*{-0.25cm}
\end{figure}

A big challenge to train this network is that there is not enough multispectral data available to yield diverse scenes.
Thus, we suggest a smart data augmentation using RGB data, since there are many large size RGB databases with thousands of different scenes.
Furthermore, realistic masks needs to be generated.

\subsubsection{Spectral Image Generation}

The main goal here is not to perfectly generate spectral data out of these images, but rather imitate that the objects have different textures for different spectral bands.
To be consistent across the whole object, the idea of this data augmentation is to map random grayscale values to a certain number of colors in the image.
This is done by transforming the RGB image to an HSV image and use the hue value to assign colors.
Between these colors, a linear interpolation is performed for consistency reasons.
Moreover, random intensities for white and black are generated to also use the saturation and value of the HSV image.
Again, linear interpolation is used.
Subsequently, the grayscale image is normalized to the full intensity range and a random exposure between 0.2 and 1.5 is applied to also cover over- and underexposed cases.
Finally, synthetic white Gaussian noise is added and the image is clipped to the intensity range.

\subsubsection{Mask Generation}

Realistic masks also need to be generated for training.
To that end, five different patterns are implemented.
First, the classic stroke mask known from inpainting papers such as \cite{nazeri_edgeconnect_2019} is used.
Moreover, a random pixel loss mask and a random block loss mask are used to simulate cases, where the depth map is noisy.
Furthermore, a border mask is provided, which just cuts off random portions of random borders of the image.
This also happens in reality, since the cameras are spatially distributed and thus cannot see all borders of the center camera.
Finally, an edge mask can be applied.
Typically, missing pixels occur on edges of objects, because often there is a bigger disparity difference between foreground and background.
Therefore, an edge mask is calculated by extracting edges from the RGB image using a Canny edge detector.
These thin edges are extended to a random width in a random direction by calculating starting and ending points of lines using the gradient map and setting all pixels on these lines to be reconstructed.
In general, more diverse masks will make the network generalize better as long as enough image content is shown to estimate the coefficients.
Examples for generated images and masks are shown in \fig\ref{fig:data_augmentation}.

\subsection{Configuration}

Since only convolutional layers are used, the network can handle any image size.
The input images need to be padded such that all cube bins cover full image areas.
Otherwise, the border linear regression coefficients will not be properly estimated.
The desired spatial bin size is set to $16 \times 16$, while there are $\gridbins=32$ luma bins.
A weighted $l_1$-loss is used as loss function
\begin{multline}
	L(\hat{\msimage}_{\channel}[\x, \y], \networkoutput_{\channel}[\x, \y], \msmask_{\channel}[\x, \y]) =\\ \frac{1}{\highwidth \highheight}\sum_{x, y} \begin{cases}
																					\alpha \cdot |\networkoutput_{\channel}[\x, \y] - \hat{\msimage}_{\channel}[\x, \y]|,& \text{if} \ \msmask_\channel[\x, \y] = 1 \\
																					|\networkoutput_{\channel}[\x, \y] - \hat{\msimage}_{\channel}[\x, \y], & \text{else},
	\end{cases}
\end{multline}
where $\alpha$ is the weight factor for missing pixel positions.
This weight factor is set to 10 to put more importance to missing regions.
Moreover, the cells in the grid should be steered towards reconstructing missing parts of the image rather than already existing parts.
However, it is still important that the whole image is reconstructed such that the network learns how images look like.
As optimizer, Adam with parameters $\beta_1 = 0.5$ and $\beta_2 = 0.999$ is picked.
The network is trained for 64 epochs using the Places2 database \cite{zhou_places_2018}, which contains approximately 1.8 million RGB images.
The learning rate starts at 0.0001 and is halved at epochs 20, 32, 40, 48, and 56.

\section{Evaluation}
\label{sec:evaluation}

The evaluation is done quantitatively as well as qualitatively.
Our novel DGNet is compared against the inpainting methods Free Form Inpainting (FF) \cite{yu_free_2019}, EdgeConnect (EC) \cite{nazeri_edgeconnect_2019}, Frequency Selective Reconstruction (FSR) \cite{genser_spectral_2018}, and the guided reconstruction methods Cross-Spectral Reconstruction (CSR) from \cite{genser_camera_2020} and Non-local Cross-Spectral Reconstruction (NOCS) \cite{sippel_spatio_2021}.

\subsection{Synthetic Data}

\begin{table}[t]
	\small
	\centering
	\caption{Average PSNR in dB and SSIM of different reconstruction methods on different databases \cite{sippel_synthetic_2023, middlebury_db_2007, genser_camera_2020}.}
	\label{tab:metrics}
	\begin{tabular}{@{}c@{\hspace*{0.1cm}}|@{\hspace*{-0.0cm}}c@{\hspace*{-0.0cm}}|@{\hspace*{0.1cm}}c@{\hspace*{0.1cm}}c@{\hspace*{0.1cm}}c@{\hspace*{0.1cm}}c@{\hspace*{0.1cm}}c@{\hspace*{0.1cm}}c@{}}
				&	Database	&	FF\cite{yu_free_2019}	&	EC\cite{nazeri_edgeconnect_2019}	&	FSR\cite{genser_spectral_2018}	&	NOCS\cite{sippel_spatio_2021}	&	CSR\cite{genser_camera_2020}	&	DGNet \\
		\hline
		\parbox[t]{2mm}{\multirow{3}{*}{\rotatebox[origin=c]{90}{PSNR}}}
				&	HyViD\cite{sippel_synthetic_2023}	&	33.72	&	29.40	&	34.32	&	35.29	&	38.98	&	\textbf{40.95}\\
				&	MiBu\cite{middlebury_db_2007}		&	28.85	&	27.38	&	30.12	&	30.94	&	33.64	&	\textbf{35.43}\\
				&	CAMSI\cite{genser_camera_2020}		&	28.57	&	28.40	&	28.21	&	29.78	&	29.26	&	\textbf{30.57}\\ \hline

		\parbox[t]{2mm}{\multirow{3}{*}{\rotatebox[origin=c]{90}{SSIM}}}
				&	HyViD\cite{sippel_synthetic_2023}	&	0.942	&	0.869	&	0.951	&	0.962	&	0.974	&	\textbf{0.989}\\
				&	MiBu\cite{middlebury_db_2007}		&	0.923	&	0.895	&	0.931	&	0.934	&	0.949	&	\textbf{0.952}\\
				&	CAMSI\cite{genser_camera_2020}		&	0.896	&	0.896	&	0.901	&	0.906	&	0.907	&	\textbf{0.913}\\
	\end{tabular}
\end{table}

\begin{table}[t]
	\small
	\centering
	\caption{Single-thread CPU runtime of all methods in seconds.}
	\label{tab:runtime}
	\vspace*{-0.5cm}
	\begin{tabular}{@{}c@{\hspace*{0.2cm}}c@{\hspace*{0.2cm}}c@{\hspace*{0.2cm}}c@{\hspace*{0.2cm}}c@{\hspace*{0.2cm}}c@{\hspace*{0.2cm}}||@{\hspace*{0.1cm}}c@{}}
		&&&&&& DGNet\\
		FF\cite{yu_free_2019}	&	EC\cite{nazeri_edgeconnect_2019}	&	FSR\cite{genser_spectral_2018}	&	NOCS\cite{sippel_spatio_2021}	&	CSR\cite{genser_camera_2020}	&	DGNet 	&	on GPU\\
		\hline
		2960	&	391		&	855		&	97.8	&	162	&	\textbf{8.27}	&	\textbf{0.34}\\
	\end{tabular}
	\vspace*{-0.2cm}
\end{table}

%

To evaluate DGNet, the HyViD database \cite{sippel_synthetic_2023} is used, which contains 7 scenes, each with 30 frames.
The images have 31 hyperspectral channels and are rendered from a camera array very similar to the one in \fig\ref{fig:problem_statement}.
Moreover, ground-truth depth data is available.
Hence, the database can be used to evaluate realistic cross spectral reconstruction problems.
For that, the 31 hyperspectral channels are synthetically filtered to nine different bandpasses, including a red, green and blue component.

The ground-truth depth is used for the reconstruction pipeline to warp the images accordingly.
In \tab\ref{tab:metrics}, the results are summarized in terms of average PSNR and SSIM, which indicates that our novel neural network-based method outperforms the reference algorithms.
Moreover, the best of three runtimes using a single thread CPU implementation is shown in \tab\ref{tab:runtime}, which are evaluated on the first frame of the scene \textit{family house}.
DGNet outperforms the reference algorithms by a factor of at least 11.8.
When executed on an RTX 3090 GPU, the runtime is decreased to 340 ms, corresponding to a factor of 287 in comparison to the second fastest method NOCS.

\begin{figure}[t]
	\begin{center}
		\begin{tikzpicture}[y=-1cm]
			\begin{scope}[shift={(0, 0)}]
				\node[inner sep=0pt, anchor=west] (rgb_2) at (0,-0.15) {\includegraphics[width=.23\textwidth, trim=0 100 0 0, clip]{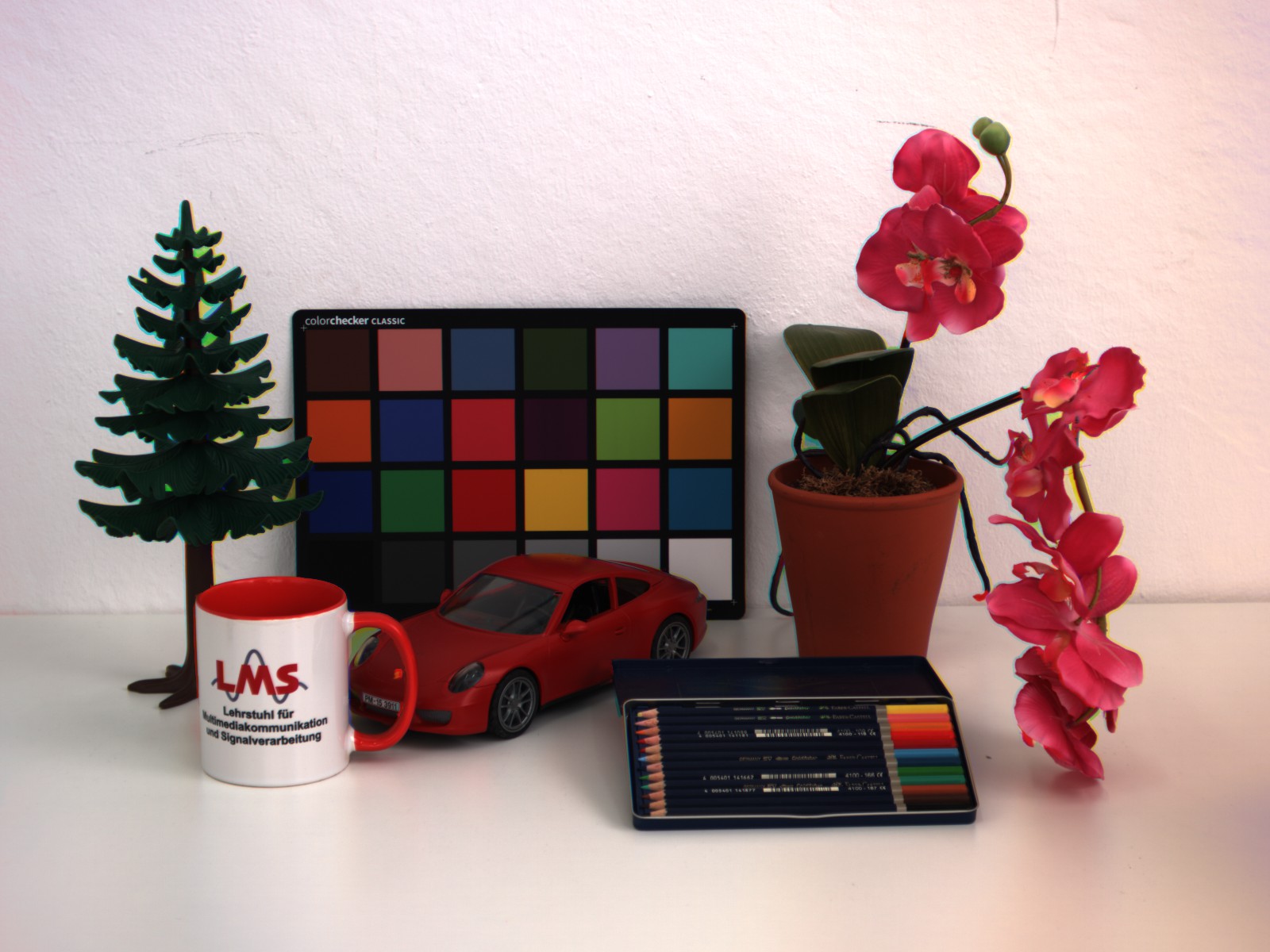}};
				\draw[draw=red, line width=0.25mm] (0.49, -0.9) rectangle ++(0.25, 0.25);
				\draw[draw=red, line width=0.25mm] (1.2,-1.5) rectangle ++(0.9, 0.9);
				\node[inner sep=0pt, anchor=west] (rgb_3) at (1.2,-1.05) {\includegraphics[width=.05\textwidth, trim=190 850 1310 250, clip]{real_nocs_345}};
				\draw[red] (0.74, -0.65) -- (1.2, -0.6);
				\draw[red] (0.74, -0.9) -- (1.2, -1.5);
				\node[fill=white, fill opacity=0.8, inner sep=1pt, anchor=west] at (0.1, -1.3) {\small NOCS};
			\end{scope}

			\begin{scope}[shift={(0, 2.8)}]
				\node[inner sep=0pt, anchor=west] (rgb_1) at (0,-0.15) {\includegraphics[width=.23\textwidth, trim=0 100 0 0, clip]{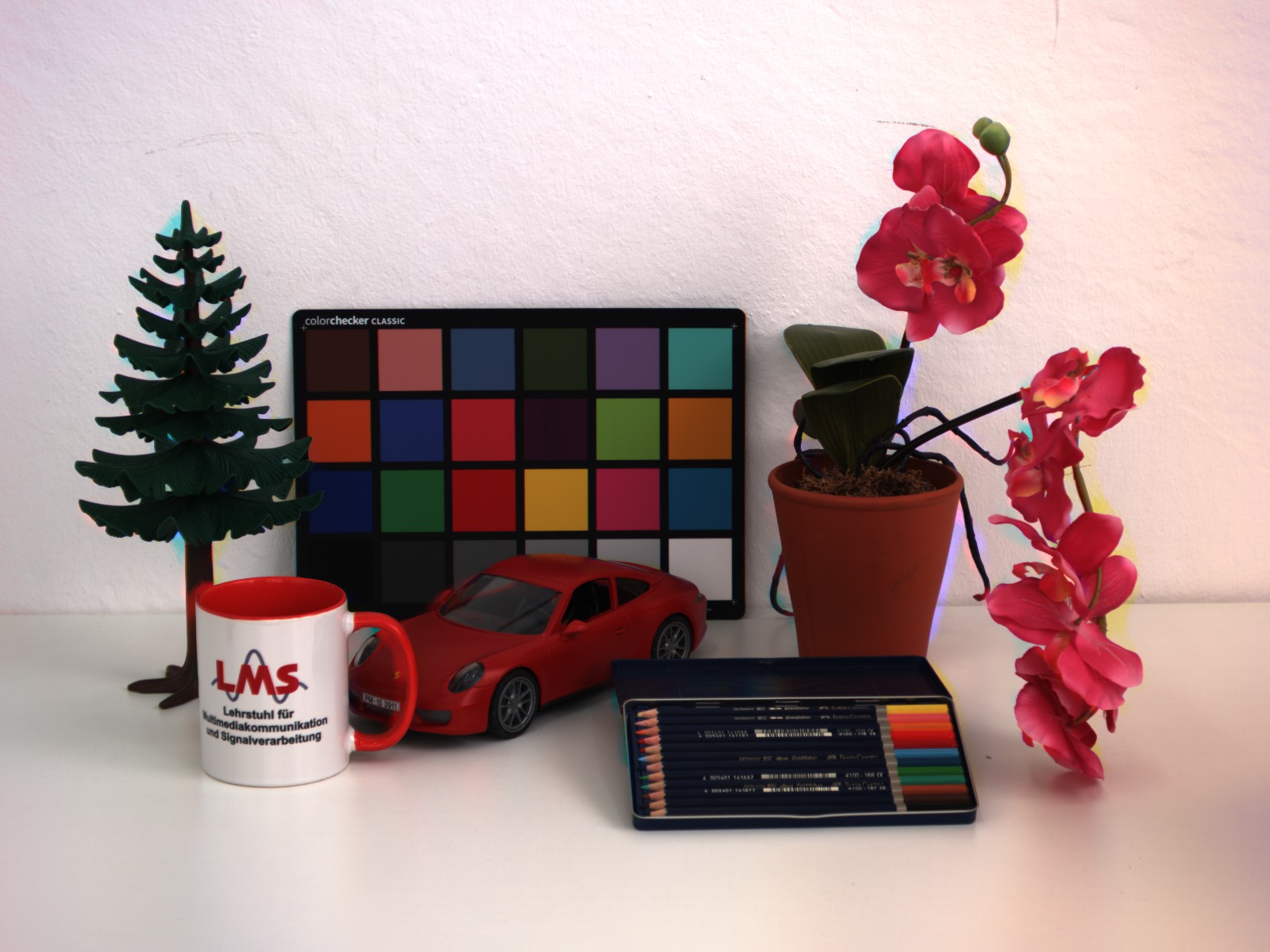}};
				\draw[draw=red, line width=0.25mm] (0.49, -0.9) rectangle ++(0.25, 0.25);
				\draw[draw=red, line width=0.25mm] (1.2, -1.5) rectangle ++(0.9, 0.9);
				\node[inner sep=0pt, anchor=west] (rgb_3) at (1.2,-1.05) {\includegraphics[width=.05\textwidth, trim=190 850 1310 250, clip]{real_matlab_345}};
				\draw[red] (0.74, -0.65) -- (1.2, -0.6);
				\draw[red] (0.74, -0.9) -- (1.2, -1.5);
				\node[fill=white, fill opacity=0.8, inner sep=1pt, anchor=west] at (0.1, -1.3) {\small CSR};
			\end{scope}

			\begin{scope}[shift={(0, 5.6)}]
				\node[inner sep=0pt, anchor=west] (rgb_3) at (0,-0.15) {\includegraphics[width=.23\textwidth, trim=0 100 0 0, clip]{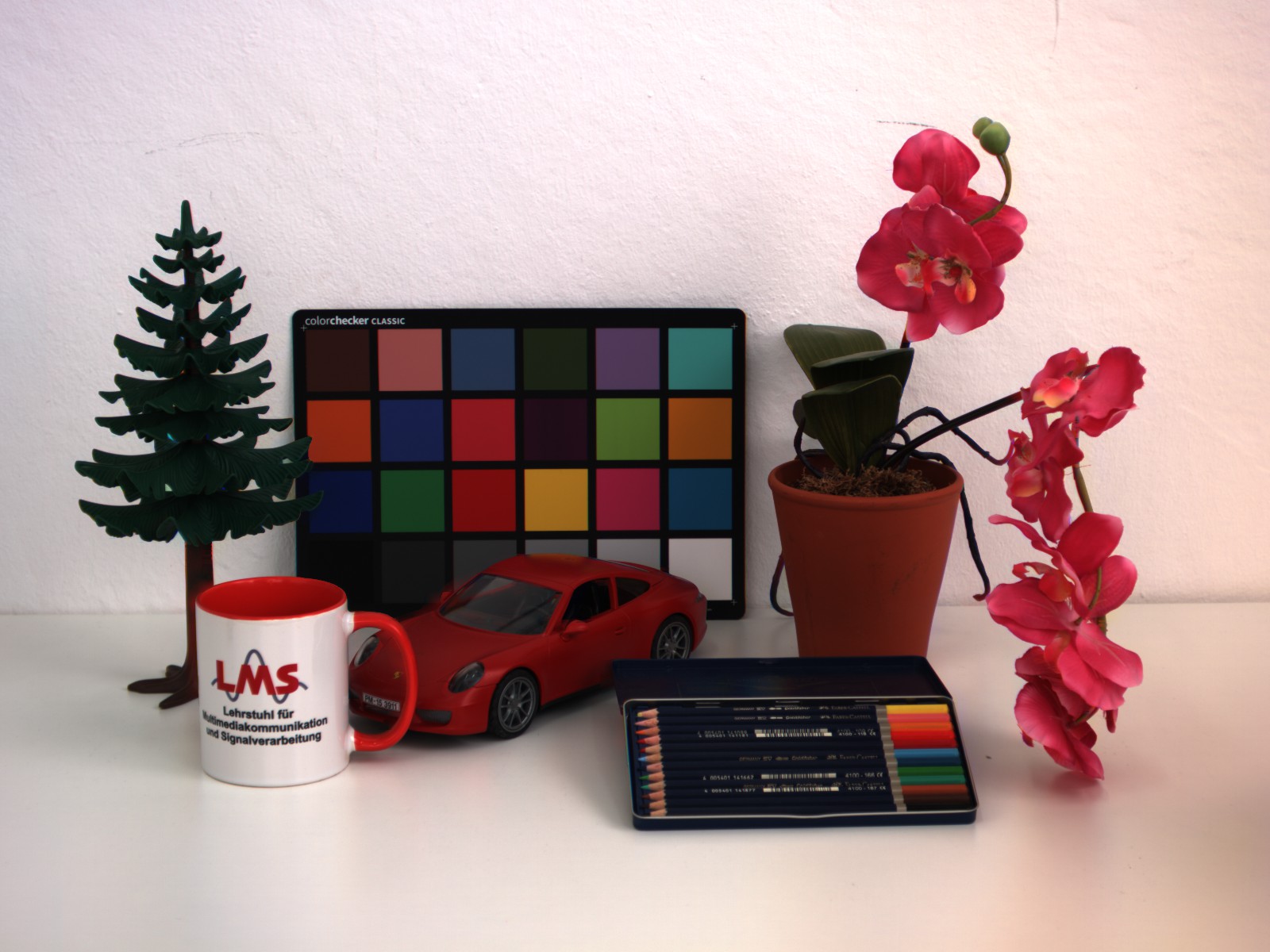}};
				\draw[draw=red, line width=0.25mm] (0.49, -0.9) rectangle ++(0.25, 0.25);
				\draw[draw=red, line width=0.25mm] (1.2, -1.5) rectangle ++(0.9, 0.9);
				\node[inner sep=0pt, anchor=west] (rgb_3) at (1.2,-1.05) {\includegraphics[width=.05\textwidth, trim=190 850 1310 250, clip]{real_dgnn_345}};
				\draw[red] (0.74, -0.65) -- (1.2, -0.6);
				\draw[red] (0.74, -0.9) -- (1.2, -1.5);
				\node[fill=white, fill opacity=0.8, inner sep=1pt, anchor=west] at (0.1, -1.3) {\small DGNet};
			\end{scope}

			\begin{scope}[shift={(4.1, 0)}]
				\node[inner sep=0pt, anchor=west] (ir_2) at (0,-0.15) {\includegraphics[width=.23\textwidth, trim=0 100 0 0, clip]{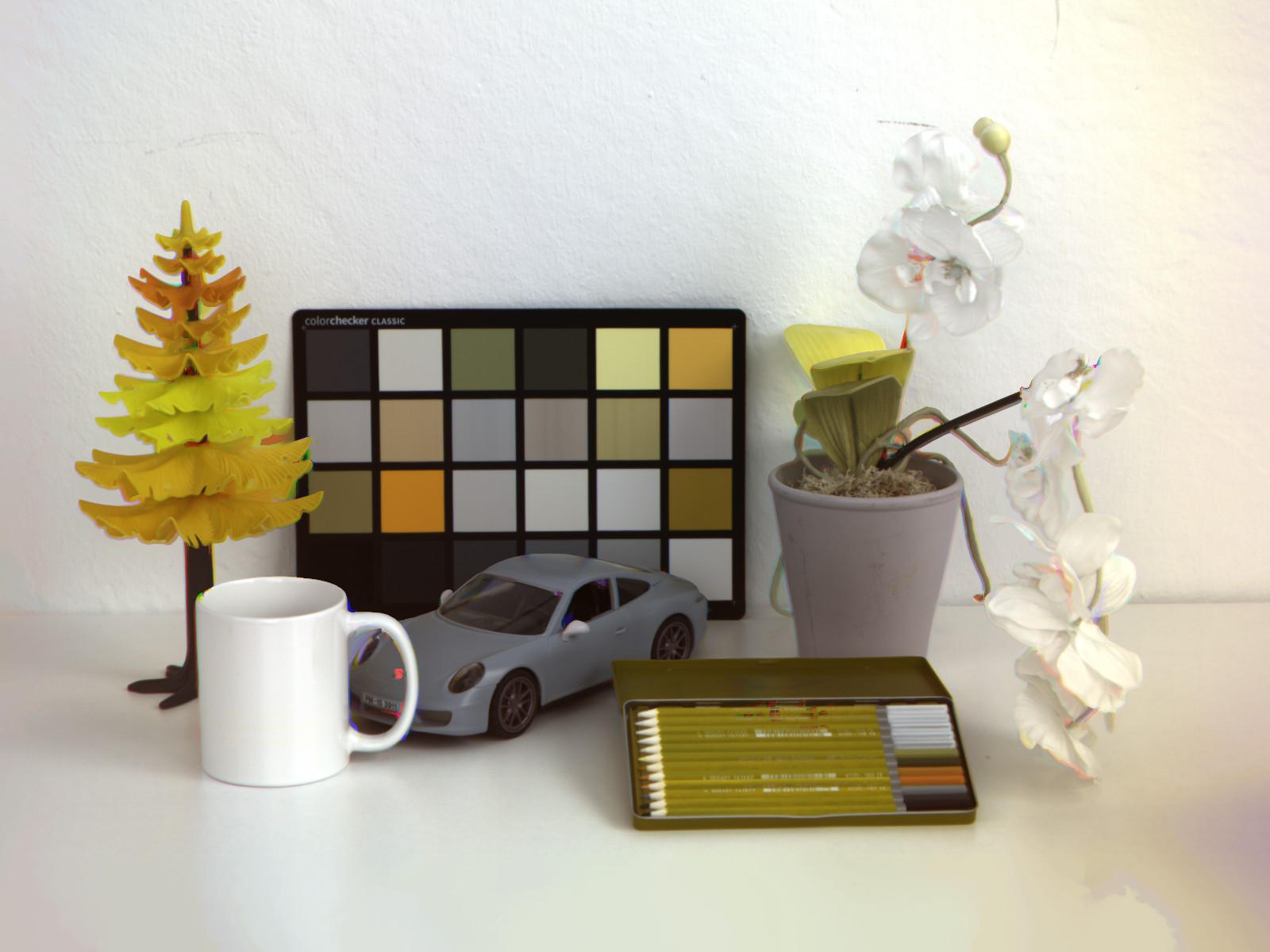}};
				\draw[draw=red, line width=0.25mm] (2.83, -0.25) rectangle ++(0.25, 0.25);
				\draw[draw=red, line width=0.25mm] (3.2,-1.5) rectangle ++(0.9, 0.9);
				\node[inner sep=0pt, anchor=west] (rgb_3) at (3.2,-1.05) {\includegraphics[width=.05\textwidth, trim=1100 600 400 500, clip]{real_nocs_012}};
				\draw[red] (3.08, 0) -- (4.1, -0.6);
				\draw[red] (2.83, -0.25) -- (3.2, -1.5);
				\node[fill=white, fill opacity=0.8, inner sep=1pt, anchor=west] at (0.1, -1.3) {\small NOCS};
			\end{scope}

			\begin{scope}[shift={(4.1, 2.8)}]
				\node[inner sep=0pt, anchor=west] (ir_1) at (0,-0.15) {\includegraphics[width=.23\textwidth, trim=0 100 0 0, clip]{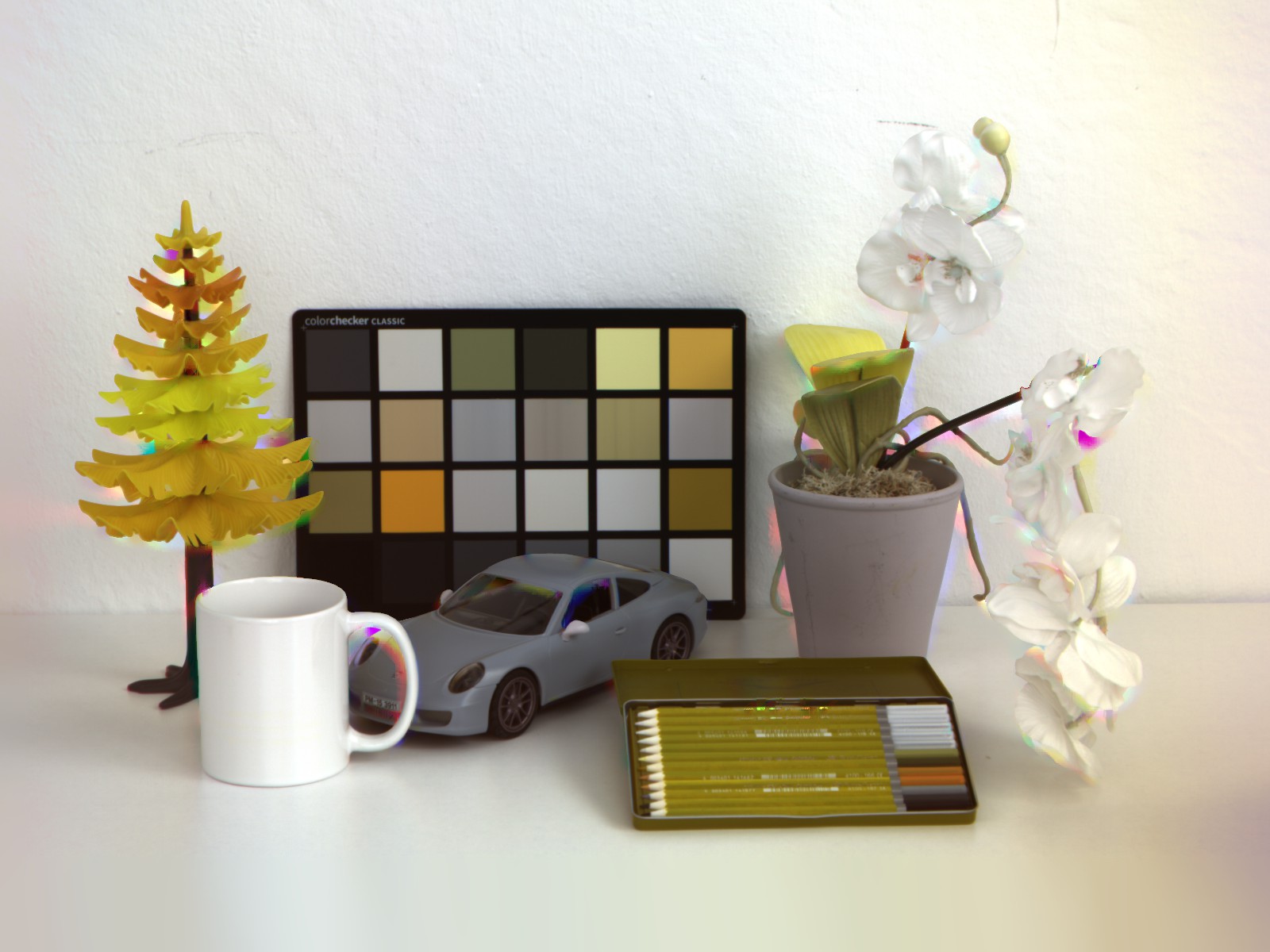}};
				\draw[draw=red, line width=0.25mm] (2.83, -0.25) rectangle ++(0.25, 0.25);
				\draw[draw=red, line width=0.25mm] (3.2,-1.5) rectangle ++(0.9, 0.9);
				\node[inner sep=0pt, anchor=west] (rgb_3) at (3.2,-1.05) {\includegraphics[width=.05\textwidth, trim=1100 600 400 500, clip]{real_matlab_012}};
				\draw[red] (3.08, 0) -- (4.1, -0.6);
				\draw[red] (2.83, -0.25) -- (3.2, -1.5);
				\node[fill=white, fill opacity=0.8, inner sep=1pt, anchor=west] at (0.1, -1.3) {\small CSR};
			\end{scope}

			\begin{scope}[shift={(4.1, 5.6)}]
				\node[inner sep=0pt, anchor=west] (ir_3) at (0,-0.15) {\includegraphics[width=.23\textwidth, trim=0 100 0 0, clip]{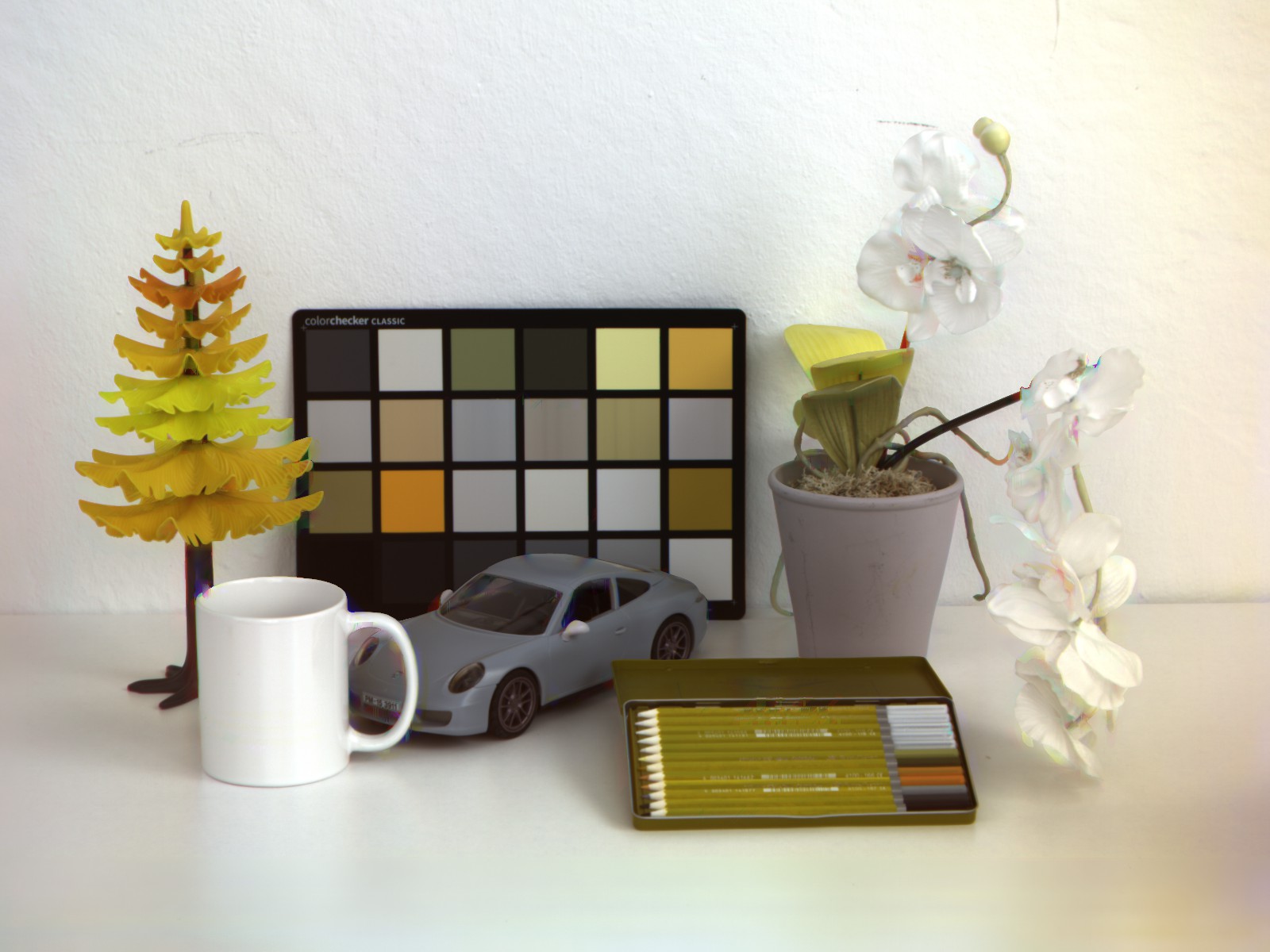}};
				\draw[draw=red, line width=0.25mm] (2.83, -0.25) rectangle ++(0.25, 0.25);
				\draw[draw=red, line width=0.25mm] (3.2,-1.5) rectangle ++(0.9, 0.9);
				\node[inner sep=0pt, anchor=west] (rgb_3) at (3.2,-1.05) {\includegraphics[width=.05\textwidth, trim=1100 600 400 500, clip]{real_dgnn_012}};
				\draw[red] (3.08, 0) -- (4.1, -0.6);
				\draw[red] (2.83, -0.25) -- (3.2, -1.5);
				\node[fill=white, fill opacity=0.8, inner sep=1pt, anchor=west] at (0.1, -1.3) {\small DGNet};
			\end{scope}
		\end{tikzpicture}
	\end{center}
	\vspace*{-0.3cm}
	\caption{Reconstruction results of spectrally guided methods as (false) color images combining three spectral bands. Left column: center row of camera array (red, green, blue). Right column: top row of camera array (infrared filters at 750 nm, 850 nm and 950 nm). Zoomed areas are depicted using red rectangles.}
	\label{fig:real_data}
	\vspace*{-0.3cm}
\end{figure}

\subsection{Real Data}

To evaluate real-world performance, an evaluation based on the Middlebury 2006 database \cite{middlebury_db_2007}, which contains 21 scenes, as well as on the database of \cite{genser_camera_2020}, which are both much smaller than the synthetic database, is shown in \tab\ref{tab:metrics}.
Note that the Middlebury database only contains RGB images.
Hence, the red channel of the left camera, the green channel of the center camera and the blue channel of the right camera are chosen to simulate a linear camera array.
Again, the proposed DGNet outperforms all reference methods.
Additionally, another image was recorded using CAMSI \cite{genser_camera_2020} for qualitative evaluation.
In \fig\ref{fig:real_data}, false color images of the cross spectral methods and our novel network are depicted.
As indicated, the novel DGNet produces fewer artefacts than its competitors.

\section{Conclusion}
\label{sec:conclusion}

This paper introduced a neural network-based method for reconstructing occluded pixels using a guide image, e.g., when employing a multispectral camera array.
For that, a cube of grid coefficients for linear regression is estimated in spatial as well as intensity direction.
Afterwards, this cube is sliced into two high-resolution linear regression parameter images using the guide image.
Finally, the linear regression coefficients are applied to this guide image to yield a final estimate of the peripheral view.
It was shown that this method outperforms the state of the art by up to 2 dB in terms of PSNR on multispectral camera array data as well as in terms of SSIM.
The novel neural network also provides a significant speedup and yields visually more accurate images.

\vfill\pagebreak
\clearpage

\bibliographystyle{IEEEbib}
\bibliography{refs}

\end{document}

%% file: figures/architecture.tex
\pgfmathsetmacro{\cubesizesmall}{0.8}
\pgfmathsetmacro{\cubesizemedium}{1.1}
\pgfmathsetmacro{\cubesizelarge}{1.4}
\pgfmathsetmacro{\cubesizehuge}{1.7}
\pgfmathsetmacro{\cubesizebiggest}{2}
\pgfmathsetmacro{\cubechannelssmall}{0.18}
\pgfmathsetmacro{\cubechannelsmedium}{0.22}
\pgfmathsetmacro{\cubechannelslarge}{0.26}
\pgfmathsetmacro{\cubechannelshuge}{0.3}
\pgfmathsetmacro{\cubechannelsbiggest}{0.34}

\node[inner sep=0pt] (lef) at (0.3,1) {\includegraphics[width=.07\textwidth]{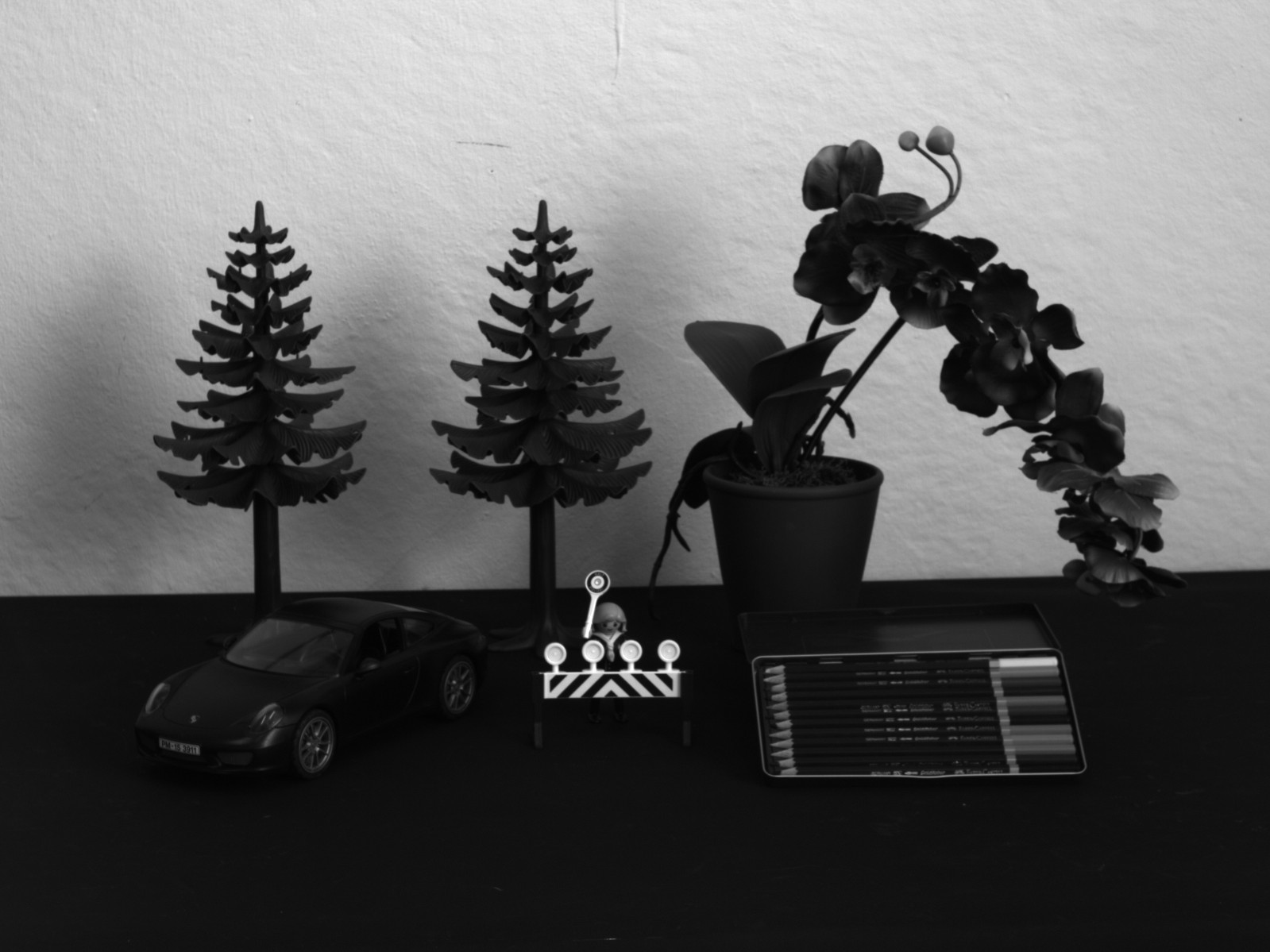}};
\node[inner sep=0pt] (lef) at (0.3,2) {\includegraphics[width=.07\textwidth]{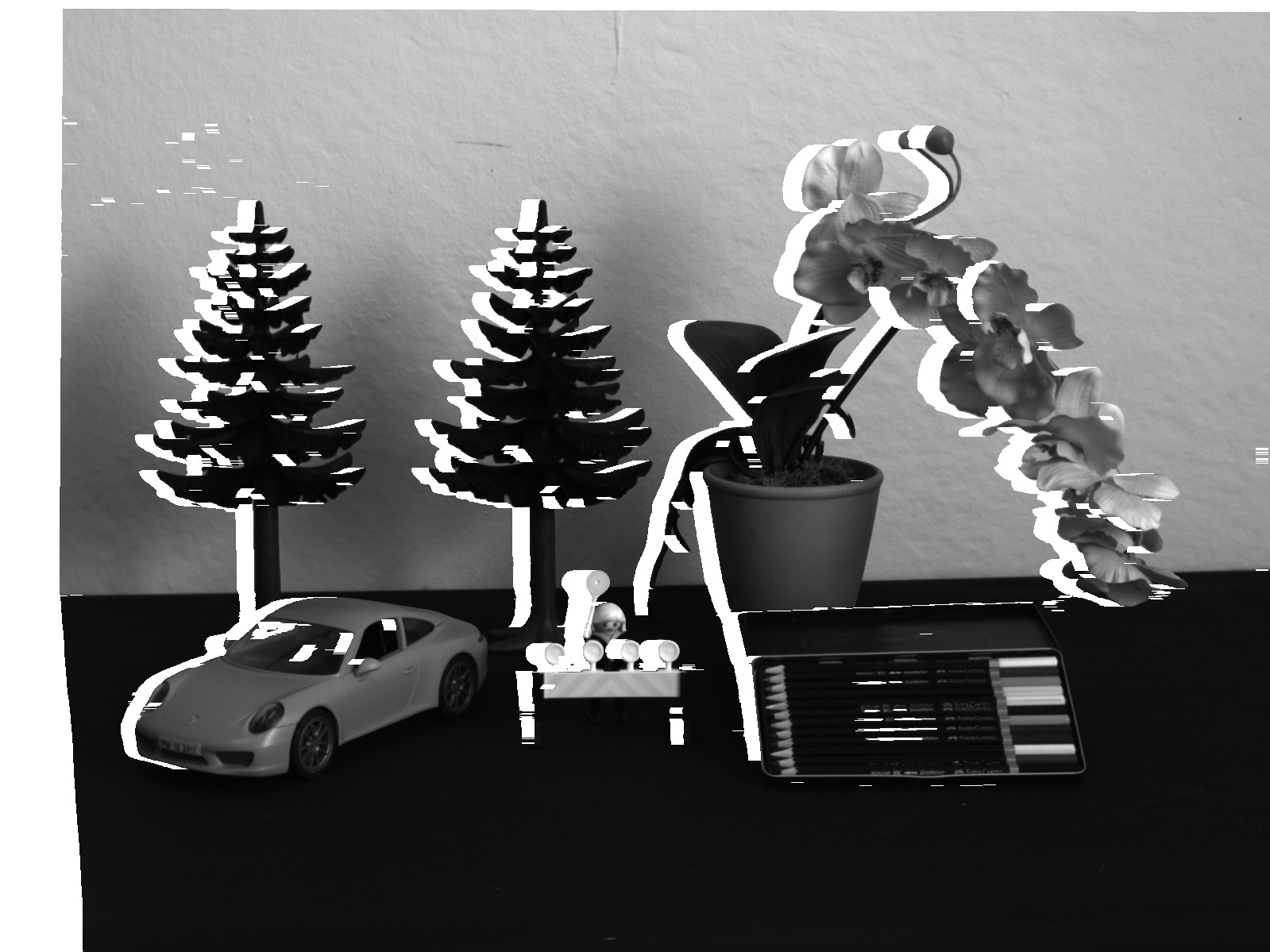}};
\node[inner sep=0pt] (lef) at (0.3,3) {\includegraphics[width=.07\textwidth]{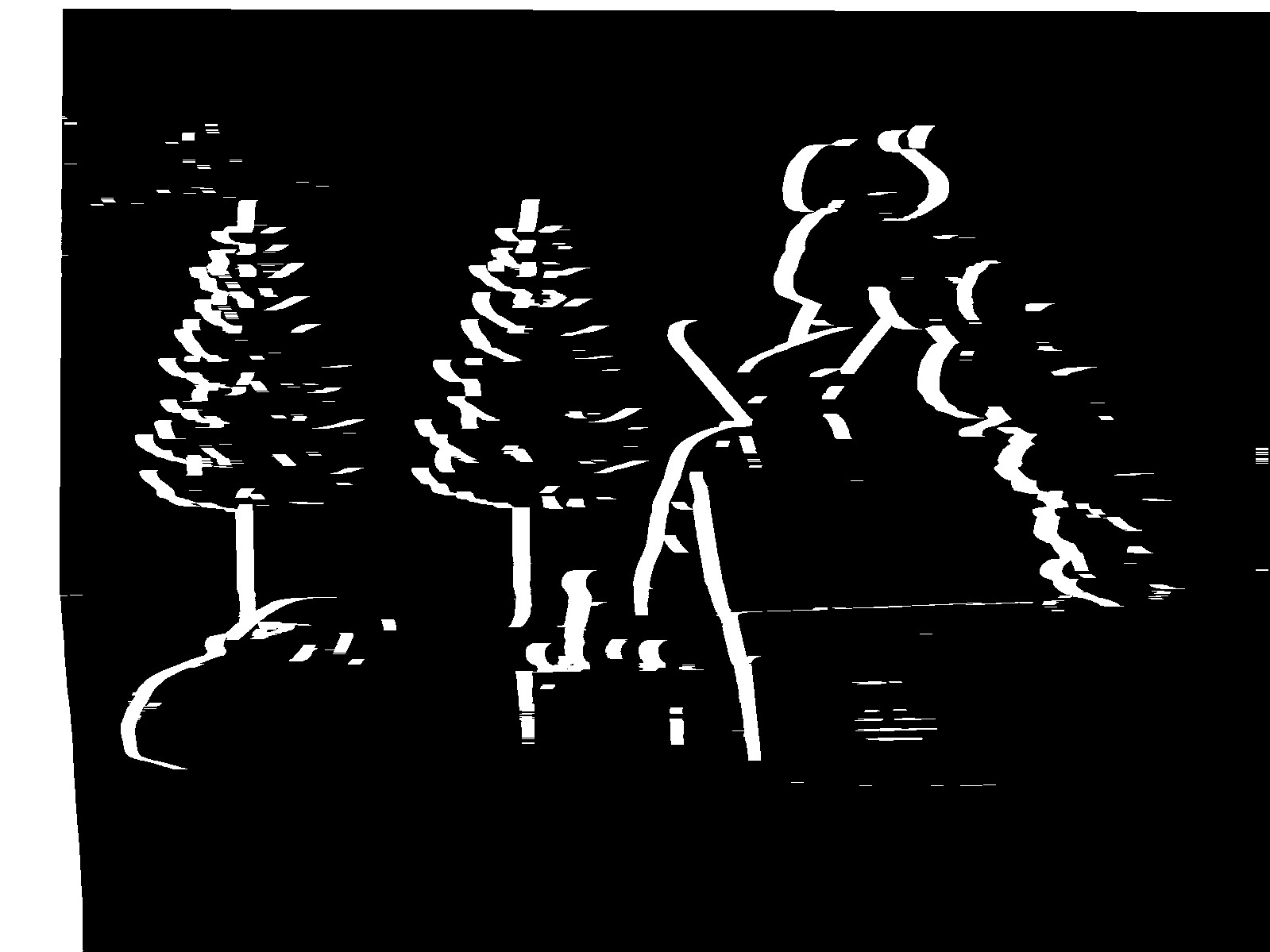}};

\node[inner sep=0pt] (lef) at (-0.9,1) {$\referenceimage[\x, \y]$};
\node[inner sep=0pt] (lef) at (-0.9,2) {$\msmasked_{\channel}[\x, \y]$};
\node[inner sep=0pt] (lef) at (-0.9,3) {$\msmask_{\channel}[\x, \y]$};

\draw[-{Stealth[scale=2]}] (1, 1) -- (2, 1.9);
\draw[-{Stealth[scale=2]}] (1, 2) -- (2, 2);
\draw[-{Stealth[scale=2]}] (1, 3) -- (2, 2.1) node[midway, below=1em] (reco) {Concat};

\begin{scope}
	[shift={(2.2, 3.5)}, x={(0:1cm)}, y={(-90:1cm)}, z={(60:0.3cm)}]
	\begin{scope}
		[shift={(0, 0)}]
		\draw[opacity=0.3, fill=cyan] (0,0,0) -- ++(\cubechannelssmall,0,0) -- ++(0,-\cubesizebiggest,0) -- ++(-\cubechannelssmall,0,0) -- cycle;
		\draw[opacity=0.3, fill=cyan] (0,-\cubesizebiggest,0) -- ++(0,0,\cubesizebiggest) -- ++(\cubechannelssmall,0,0) -- ++(0,0,-\cubesizebiggest) -- cycle;
		\draw[opacity=0.3, fill=cyan] (\cubechannelssmall,-\cubesizebiggest,0) -- ++(0,0,\cubesizebiggest) -- ++(0,\cubesizebiggest,0) -- ++(0,0,-\cubesizebiggest) -- cycle;

  		\node [rotate=90, anchor=west] at (0.1, 0.1) {\tiny 32/3/2\textdownarrow};

	\end{scope}
	\begin{scope}
		[shift={(0.6, 0)}]
		\draw[opacity=0.3, fill=cyan] (0,0,0) -- ++(\cubechannelsmedium,0,0) -- ++(0,-\cubesizehuge,0) -- ++(-\cubechannelsmedium,0,0) -- cycle;
		\draw[opacity=0.3, fill=cyan] (0,-\cubesizehuge,0) -- ++(0,0,\cubesizehuge) -- ++(\cubechannelsmedium,0,0) -- ++(0,0,-\cubesizehuge) -- cycle;
		\draw[opacity=0.3, fill=cyan] (\cubechannelsmedium,-\cubesizehuge,0) -- ++(0,0,\cubesizehuge) -- ++(0,\cubesizehuge,0) -- ++(0,0,-\cubesizehuge) -- cycle;

  		\node [rotate=90, anchor=west] at (0.11, 0.1) {\tiny 64/3/2\textdownarrow};
	\end{scope}
	\begin{scope}
		[shift={(1.15, 0)}]
		\draw[opacity=0.3, fill=cyan] (0,0,0) -- ++(\cubechannelslarge,0,0) -- ++(0,-\cubesizelarge,0) -- ++(-\cubechannelslarge,0,0) -- cycle;
		\draw[opacity=0.3, fill=cyan] (0,-\cubesizelarge,0) -- ++(0,0,\cubesizelarge) -- ++(\cubechannelslarge,0,0) -- ++(0,0,-\cubesizelarge) -- cycle;
		\draw[opacity=0.3, fill=cyan] (\cubechannelslarge,-\cubesizelarge,0) -- ++(0,0,\cubesizelarge) -- ++(0,\cubesizelarge,0) -- ++(0,0,-\cubesizelarge) -- cycle;

  		\node [rotate=90, anchor=west] at (0.12, 0.1) {\tiny 128/3/2\textdownarrow};
	\end{scope}
	\begin{scope}
		[shift={(1.7, 0)}]
		\draw[opacity=0.3, fill=cyan] (0,0,0) -- ++(\cubechannelshuge,0,0) -- ++(0,-\cubesizemedium,0) -- ++(-\cubechannelshuge,0,0) -- cycle;
		\draw[opacity=0.3, fill=cyan] (0,-\cubesizemedium,0) -- ++(0,0,\cubesizemedium) -- ++(\cubechannelshuge,0,0) -- ++(0,0,-\cubesizemedium) -- cycle;
		\draw[opacity=0.3, fill=cyan] (\cubechannelshuge,-\cubesizemedium,0) -- ++(0,0,\cubesizemedium) -- ++(0,\cubesizemedium,0) -- ++(0,0,-\cubesizemedium) -- cycle;

  		\node [rotate=90, anchor=west] at (0.13, 0.1) {\tiny 256/3/2\textdownarrow};
	\end{scope}
	\foreach \x in{0,...,7}
		{
		\begin{scope}
			[shift={(2.2 + \x * 0.5, 0)}]
			\draw[opacity=0.3, fill=cyan] (0,0,0) -- ++(\cubechannelsbiggest,0,0) -- ++(0,-\cubesizesmall,0) -- ++(-\cubechannelsbiggest,0,0) -- cycle;
			\draw[opacity=0.3, fill=cyan] (0,-\cubesizesmall,0) -- ++(0,0,\cubesizesmall) -- ++(\cubechannelsbiggest,0,0) -- ++(0,0,-\cubesizesmall) -- cycle;
			\draw[opacity=0.3, fill=cyan] (\cubechannelsbiggest,-\cubesizesmall,0) -- ++(0,0,\cubesizesmall) -- ++(0,\cubesizesmall,0) -- ++(0,0,-\cubesizesmall) -- cycle;

  			\node [rotate=90, anchor=west] at (0.14, 0.1) {\tiny 256/3/1};
		\end{scope}
	}

	\begin{scope}
		[shift={(6.2, 0)}]
		\draw[opacity=0.3, fill=cyan] (0,0,0) -- ++(\cubechannelsmedium,0,0) -- ++(0,-\cubesizesmall,0) -- ++(-\cubechannelsmedium,0,0) -- cycle;
		\draw[opacity=0.3, fill=cyan] (0,-\cubesizesmall,0) -- ++(0,0,\cubesizesmall) -- ++(\cubechannelsmedium,0,0) -- ++(0,0,-\cubesizesmall) -- cycle;
		\draw[opacity=0.3, fill=cyan] (\cubechannelsmedium,-\cubesizesmall,0) -- ++(0,0,\cubesizesmall) -- ++(0,\cubesizesmall,0) -- ++(0,0,-\cubesizesmall) -- cycle;

  		\node [rotate=90, anchor=west] at (0.11, 0.1) {\tiny 64/3/1};
	\end{scope}

	\begin{scope}
		[shift={(6.8, 0)}, scale=0.25]
		\newcommand{\densityx}{1}
		\newcommand{\densityy}{1}
		\newcommand{\densityz}{1}
		\pgfmathsetmacro{\steppingx}{1/\densityx}
		\pgfmathsetmacro{\steppingy}{1/\densityy}
		\pgfmathsetmacro{\steppingz}{1/\densityz}
		\newcommand{\dimx}{4}
		\newcommand{\dimy}{4}
		\newcommand{\dimz}{4}
		\pgfmathsetmacro{\secondx}{2*\steppingx}
		\pgfmathsetmacro{\secondy}{2*\steppingy}
		\pgfmathsetmacro{\secondz}{2*\steppingz}
		\foreach \x in {\steppingx,\secondx,...,\dimx}
		{
			\foreach \y in {\steppingy,\secondy,...,\dimy}
			{
				\pgfmathsetmacro{\lowx}{(\x-\steppingx)}
				\pgfmathsetmacro{\lowy}{(-\y+\steppingy)}
				\filldraw[fill=yellow,draw=black,opacity=0.3] (\lowx,\lowy,0) -- (\lowx,-\y,0) -- (\x,-\y,0) -- (\x,\lowy,0) -- cycle;

			}
		}
		\foreach \x in {\steppingx,\secondx,...,\dimx}
		{
			\foreach \z in {\steppingz,\secondz,...,\dimz}
			{
				\pgfmathsetmacro{\lowx}{(\x-\steppingx)}
				\pgfmathsetmacro{\lowz}{(\z-\steppingz)}
				\filldraw[fill=yellow,draw=black,opacity=0.3] (\lowx,-\dimy,\lowz) -- (\lowx,-\dimy,\z) -- (\x,-\dimy,\z) -- (\x,-\dimy,\lowz) -- cycle;
			}
		}
		\foreach \y in {\steppingy,\secondy,...,\dimy}
		{
			\foreach \z in {\steppingz,\secondz,...,\dimz}
			{
				\pgfmathsetmacro{\lowy}{(-\y+\steppingy)}
				\pgfmathsetmacro{\lowz}{(\z-\steppingz)}
				\filldraw[fill=yellow,draw=black,opacity=0.3] (\dimx,\lowy,\lowz) -- (\dimx,\lowy,\z) -- (\dimx,-\y,\z) -- (\dimx,-\y,\lowz) -- cycle;
			}
		}
	\end{scope}
\end{scope}

\draw[-{Stealth[scale=2]}] (1, 0.5) -- (13.4, 0.5) node[midway, below] (guide) {Guide image};
\draw[-{Stealth[scale=2]}] (9.5, 1.7) -- (11, 1.7) node[midway, above=0.2em] (grid) {Grid coefficients};

\begin{scope}
	[shift={(12, 1.8)}]
	\node[inner sep=0pt] (lef) at (0.5,1) {\includegraphics[width=.1\textwidth]{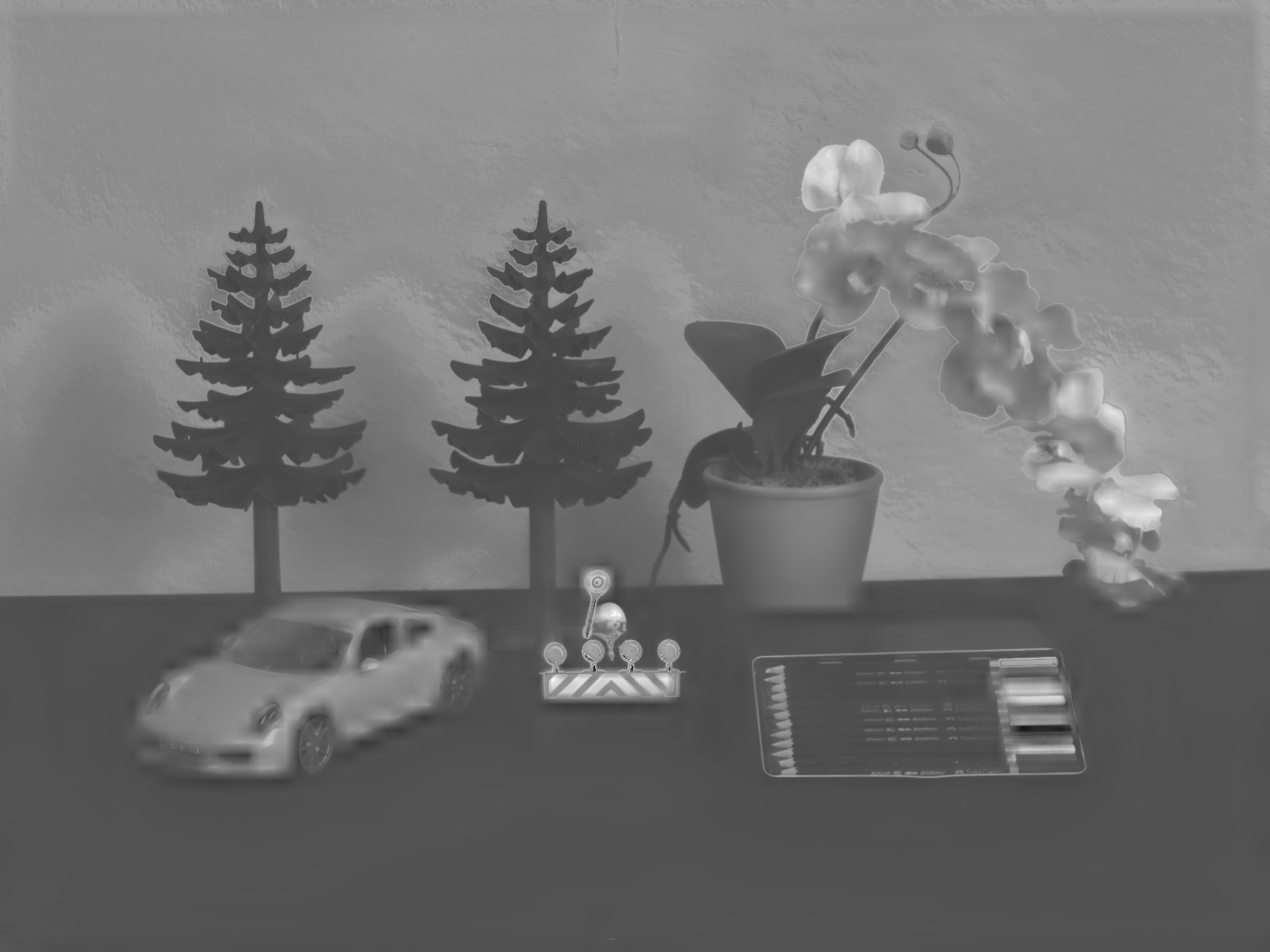}};
	\node[inner sep=0pt] (lef) at (0,0.5) {\includegraphics[width=.1\textwidth]{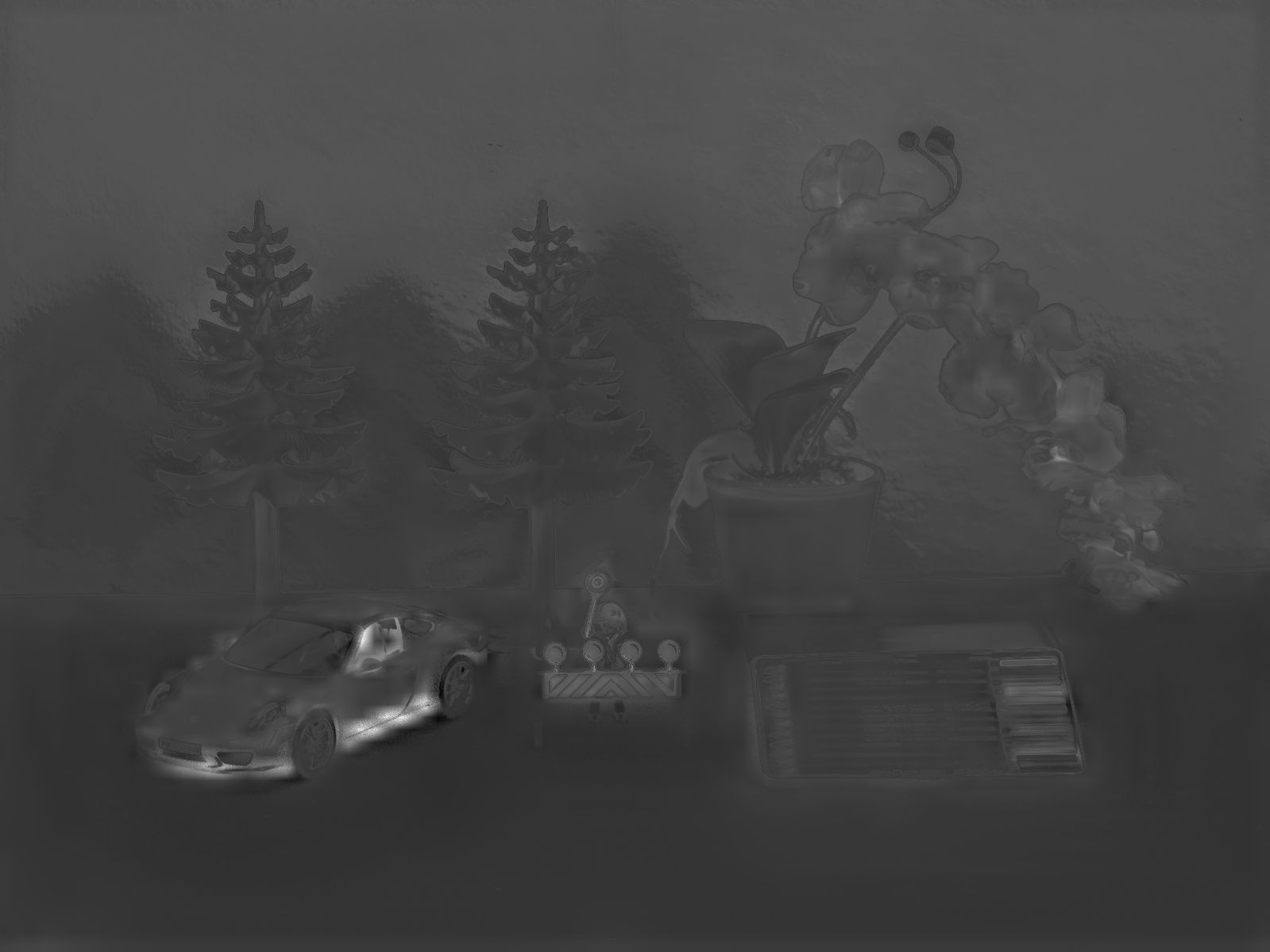}};
\end{scope}

\draw[-{Stealth[scale=2]}] (13.6, 3) -| (14.5, 2.5) node[midway, below] (grid) {LR parameters};

\draw[-{Stealth[scale=2]}] (12, 0.5) -- (12, 1.5);

\node[inner sep=0pt] (lef) at (14.9,1.4) {\includegraphics[width=.15\textwidth]{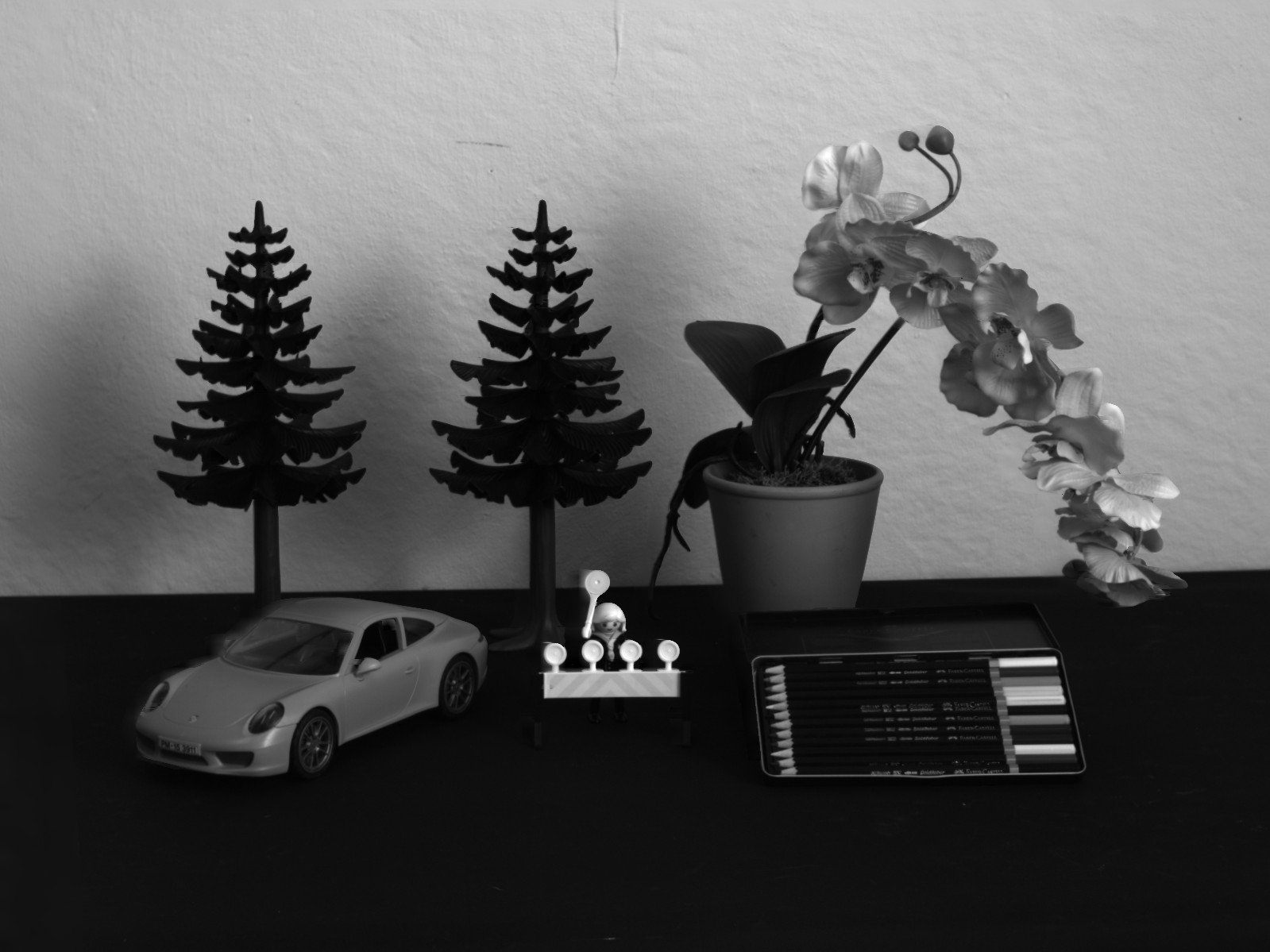}};
\node[inner sep=0pt] (lef) at (15.7,2.6) {$\networkoutput_{\channel}[\x, \y]$};

%% file: paper.bbl
\begin{thebibliography}{10}

\bibitem{cam_array_wheat_2021}
S.~Dandrifosse, A.~Carlier, B.~Dumont, and B.~Mercatoris,
\newblock ``Registration and fusion of close-range multimodal wheat images in
  field conditions,''
\newblock {\em Remote Sensing}, vol. 13, no. 7, 2021.

\bibitem{cam_array_notch_2022}
F.~Huang, P.~Lin, R.~Cao, B.~Zhou, and X.~Wu,
\newblock ``Dictionary learning- and total variation-based
  high-light-efficiency snapshot multi-aperture spectral imaging,''
\newblock {\em Remote Sensing}, vol. 14, no. 16, 2022.

\bibitem{cam_array_art_2022}
Á. Gómez~Manzanares, D.~Vázquez~Moliní, A.~Alvarez Fernandez-Balbuena,
  S.~Mayorga~Pinilla, and J.~C. Martínez~Antón,
\newblock ``Measuring high dynamic range spectral reflectance of artworks
  through an image capture matrix hyperspectral camera,''
\newblock {\em Sensors}, vol. 22, no. 13, 2022.

\bibitem{moroni_pet_2015}
M.~Moroni, A.~Mei, A.~Leonardi, E.~Lupo, and F.~Marca,
\newblock ``{PET} and {PVC} {Separation} with {Hyperspectral} {Imagery},''
\newblock {\em Sensors}, vol. 15, no. 1, pp. 2205--2227, Jan. 2015.

\bibitem{edelman_hyperspectral_2013}
G.~J. Edelman, T.~G. van Leeuwen, and M.~C.~G. Aalders,
\newblock ``{Hyperspectral imaging of the crime scene for detection and
  identification of blood stains},''
\newblock in {\em Algorithms and Technologies for Multispectral, Hyperspectral,
  and Ultraspectral Imagery XIX}, Sylvia~S. Shen and Paul~E. Lewis, Eds.
  International Society for Optics and Photonics, 2013, vol. 8743, p. 87430A,
  SPIE.

\bibitem{lima_monitoring_2020}
M.~Cardim Ferreira~Lima, A.~Krus, C.~Valero, A.~Barrientos, J.~del Cerro, and
  J.~J. Roldán-Gómez,
\newblock ``Monitoring plant status and fertilization strategy through
  multispectral images,''
\newblock {\em Sensors}, vol. 20, no. 2, 2020.

\bibitem{genser_camera_2020}
N.~Genser, J.~Seiler, and A.~Kaup,
\newblock ``Camera {Array} for {Multi}-{Spectral} {Imaging},''
\newblock {\em IEEE Transactions on Image Processing}, vol. 29, pp. 9234--9249,
  2020.

\bibitem{genser_deep_2020}
N.~Genser, A.~Spruck, J.~Seiler, and A.~Kaup,
\newblock ``Deep learning based cross-spectral disparity estimation for stereo
  imaging,''
\newblock in {\em Proc. IEEE International Conference on Image Processing
  (ICIP)}, 2020, pp. 2536--2540.

\bibitem{getreuer_total_2012}
P.~Getreuer,
\newblock ``Total {Variation} {Inpainting} using {Split} {Bregman},''
\newblock {\em Image Processing On Line}, vol. 2, pp. 147--157, July 2012.

\bibitem{genser_spectral_2018}
N.~Genser, J.~Seiler, and A.~Kaup,
\newblock ``Spectral {Constrained} {Frequency} {Selective} {Extrapolation} for
  {Rapid} {Image} {Error} {Concealment},''
\newblock in {\em Proc. 25th {International} {Conference} on {Systems},
  {Signals} and {Image} {Processing} ({IWSSIP})}. June 2018, pp. 1--5, IEEE.

\bibitem{nazeri_edgeconnect_2019}
K.~Nazeri, E.~Ng, T.~Joseph, F.~Qureshi, and M.~Ebrahimi,
\newblock ``Edgeconnect: Structure guided image inpainting using edge
  prediction,''
\newblock in {\em Proc. IEEE/CVF International Conference on Computer Vision
  Workshop (ICCVW)}, 2019, pp. 3265--3274.

\bibitem{yu_free_2019}
J.~Yu, Z.~Lin, J.~Yang, X.~Shen, X.~Lu, and T.~Huang,
\newblock ``Free-form image inpainting with gated convolution,''
\newblock in {\em Proc. IEEE/CVF International Conference on Computer Vision
  (ICCV)}, 2019, pp. 4470--4479.

\bibitem{liu_image_2018}
G.~Liu, F.~A. Reda, K.~J. Shih, T.~Wang, A.~Tao, and B.~Catanzaro,
\newblock ``Image inpainting for irregular holes using partial convolutions,''
\newblock in {\em Proc. European Conference on Computer Vision (ECCV)}, Sept.
  2018, pp. 85--100.

\bibitem{sippel_spatio_2021}
F.~Sippel, J.~Seiler, and A.~Kaup,
\newblock ``Spatio-spectral image reconstruction using non-local filtering,''
\newblock in {\em Proc. International Conference on Visual Communications and
  Image Processing (VCIP)}, 2021, pp. 1--5.

\bibitem{bilateral_chen_2016}
J.~Chen, A.~Adams, N.~Wadhwa, and S.~W. Hasinoff,
\newblock ``Bilateral guided upsampling,''
\newblock {\em ACM Trans. Graph.}, vol. 35, no. 6, Dec. 2016.

\bibitem{gharbi_deep_2017}
M.~Gharbi, J.~Chen, J.~T. Barron, S.~W. Hasinoff, and F.~Durand,
\newblock ``Deep bilateral learning for real-time image enhancement,''
\newblock {\em ACM Trans. Graph.}, vol. 36, no. 4, July 2017.

\bibitem{zhou_places_2018}
B.~Zhou, A.~Lapedriza, A.~Khosla, A.~Oliva, and A.~Torralba,
\newblock ``Places: A 10 million image database for scene recognition,''
\newblock {\em IEEE Transactions on Pattern Analysis and Machine Intelligence},
  vol. 40, no. 6, pp. 1452--1464, 2018.

\bibitem{sippel_synthetic_2023}
F.~Sippel, J.~Seiler, and A.~Kaup,
\newblock ``Synthetic hyperspectral array video database with applications to
  cross-spectral reconstruction and hyperspectral video coding,''
\newblock {\em J. Opt. Soc. Am. A}, vol. 40, no. 3, pp. 479--491, Mar. 2023.

\bibitem{middlebury_db_2007}
D.~Scharstein and C.~Pal,
\newblock ``Learning conditional random fields for stereo,''
\newblock in {\em Proc. IEEE Conference on Computer Vision and Pattern
  Recognition}, 2007, pp. 1--8.

\end{thebibliography}
